\documentclass[fleqn,12pt]{article}
\usepackage{amsfonts,epsfig,latexsym}
\usepackage[vcentermath,enableskew]{youngtab}

\makeatletter
\newbox\slashbox \setbox\slashbox=\hbox{$/$}
\def\pFMslash#1{\setbox\@tempboxa=\hbox{$#1$}
  \@tempdima=0.5\wd\slashbox \advance\@tempdima 0.5\wd\@tempboxa
  \copy\slashbox \kern-\@tempdima \box\@tempboxa}
\newcommand\FMslash{\protect\pFMslash}
\makeatother

\newcommand{\Ga}{\alpha}
\newcommand{\Gb}{\beta}

\newcommand{\Gd}{\delta}
\newcommand{\Ge}{\epsilon}
\newcommand{\Geps}{\varepsilon}
\newcommand{\Gg}{\gamma}
\newcommand{\GG}{\Gamma}

\newcommand{\Gth}{\theta}
\newcommand{\GTh}{\Theta}

\newcommand{\cA}{{\scriptscriptstyle\cal A}}
\newcommand{\cB}{{\scriptscriptstyle\cal B}}

\newcommand{\cM}{{\scriptscriptstyle\cal M}}
\newcommand{\cN}{{\scriptscriptstyle\cal N}}
\newcommand{\cK}{{\scriptscriptstyle\cal K}}
\newcommand{\cL}{{\scriptscriptstyle\cal L}}
\newcommand{\cP}{{\scriptscriptstyle\cal P}}
\newcommand{\cQ}{{\scriptscriptstyle\cal Q}}
\newcommand{\cR}{{\scriptscriptstyle\cal R}}

\newcommand{\CD}{{\cal D}}

\newcommand{\CM}{{\cal M}}

\newcommand{\CK}{{\cal K}}
\newcommand{\CL}{{\cal L}}
\newcommand{\CP}{{\cal P}}

\newcommand{\CS}{{\cal S}}
\newcommand{\CT}{{\cal T}}

\newcommand{\CV}{{\cal V}}

\newcommand{\dA}{{\dot{A}}}

\newcommand{\Be}{{\bar{\Ge}{}}}
\newcommand{\Bchi}{{\bar{\chi}{}}}
\newcommand{\Bpsi}{{\bar{\psi}{}}}
\newcommand{\Bphi}{{\bar{\phi}{}}}

\newcommand{\Bi}{{\bar{\imath}}}
\newcommand{\Bj}{{\bar{\jmath}}}
\newcommand{\eqn}[1]{(\ref{#1})}

\newcommand{\bbP}{\mathbb{P}}
\newcommand{\bbT}{\Pi}

\newcommand{\ft}[2]{{\textstyle {\frac{#1}{#2}} }}
\newcommand{\dd}{\partial}
\newcommand{\tr}{{\rm tr \,}}

\newcommand{\I}{{\rm i}}

\newcommand{\be}{\begin{equation}}
\newcommand{\ee}{\end{equation}}
\newcommand{\ben}{\begin{displaymath}}
\newcommand{\een}{\end{displaymath}}
\newcommand{\ba}{\begin{eqnarray}}
\newcommand{\ea}{\end{eqnarray}}
\newcommand{\nn}{\nonumber}
\newcommand{\non}{\nonumber\\}
\newcommand{\bean}{\begin{eqnarray*}}
\newcommand{\eean}{\end{eqnarray*}}

\newcommand{\mathon}{\mathversion{bold}}
\newcommand{\mathoff}{\mathversion{normal}}

\def\moth{\mathsurround=0pt}
\newdimen\zo \zo=0pt

\def\tick{\leaders\hrule height 0.5ex depth 0pt \hskip 0.5pt}
\def\upboxfill{$\moth \setbox\zo\hbox{\tick}%
  \hskip 2pt\hbox to 0pt{$\tick$\hss}\hrulefill \hbox to
6pt{$\tick$\hss}$}
\def\underbox#1{\offinterlineskip{\mathord{\mathop{\vtop{\moth\ialign{##\crcr
      $\hfil\displaystyle{#1}\hfil$\crcr\noalign{}
      {\upboxfill}\crcr\noalign{}}}}\limits}}}
\def\dtick{\leaders\hrule height .34pt depth .5ex \hskip 0.5pt}
\def\downboxfill{$\moth \setbox\zo\hbox{\dtick}%
  \hskip 2pt\hbox to 0pt{$\dtick$\hss}\hrulefill \hbox to
6pt{$\dtick$\hss}$}
\def\overbox#1{\mathop{\vbox{\moth\ialign{##\crcr\noalign{}
\downboxfill\crcr\noalign{\vskip 1pt\nointerlineskip}
      $\hfil\displaystyle{#1}\hfil$\crcr}}}\limits}

\def\undersym#1{\underbox{{}#1}}
\def\oversym#1{\overbox{{}#1}}

\newcommand{\Ref}[1]{(\ref{#1})}

\makeatletter
\@addtoreset{equation}{section}
\makeatother

\newcommand{\EE}[2]{{{\rm E}_{{#1}({#2})}}}
\newcommand{\EF}[2]{{{\rm F}_{{#1}({#2})}}}
\newcommand{\EG}[2]{{{\rm G}_{{#1}({#2})}}}
\newcommand{\SL}[1]{{{\rm SL}({#1})}}
\newcommand{\SU}[1]{{{\rm SU}({#1})}}
\newcommand{\SO}[1]{{{\rm SO}({#1})}}
\newcommand{\Sp}[1]{{{\rm Sp}({#1})}}

\newcommand{\UU}[1]{{{\rm U}({#1})}}

\newcommand{\vl}{{\vphantom{[}}}

\newcommand{\cro}{\!\times\!}

\newcommand{\pls}{\!+\!}
\newcommand{\mis}{\!-\!}

\newcommand{\Adj}{R_{\rm adj}}

\newcommand{\Mat}{L}

\begin{document}

\thispagestyle{empty}

\begin{flushright}
ITP-UU-03/25\\
SPIN-03/16
\end{flushright}

\smallskip

\begin{center}
\mathon
{\bf\large GAUGED LOCALLY SUPERSYMMETRIC \\ [4mm]
${D}\!=\!3$ NONLINEAR SIGMA MODELS}
\mathoff

\bigskip\bigskip

{\bf Bernard de Wit, Ivan Herger and Henning Samtleben } 

\vspace{.3cm}

{\em Institute for Theoretical Physics}  \& 
{\em Spinoza Institute,\\[.5ex]
Utrecht University, Postbus 80.195, 3508 TD Utrecht, 
The Netherlands
}
\smallskip

{\small
{\tt B.deWit@phys.uu.nl},
{\tt I.Herger@phys.uu.nl},
{\tt H.Samtleben@phys.uu.nl}
}

\end{center}

\renewcommand{\thefootnote}{\arabic{footnote}}
\setcounter{footnote}{0}
\bigskip
\medskip
\begin{abstract}
We construct supersymmetric deformations of general, locally
supersymmetric, nonlinear sigma models in three spacetime dimensions,
by extending the pure supergravity theory with a Chern-Simons term and
gauging a subgroup of the sigma model isometries, possibly augmented
with R-symmetry transformations. This class of models is shown to
include theories with standard Yang-Mills Lagrangians, with optional
moment interactions and topological mass terms. The results constitute
a general classification of three-dimensional gauged supergravities.
\end{abstract}
\vfill

\newpage

\tableofcontents

\mathon
\section{Introduction}
\mathoff
Supergravity theories with vector gauge fields can usually be
deformed by introducing gauge charges for the various
fields. These charges generate a corresponding gauge
group. Supersymmetry then necessitates the presence of additional
interactions consisting of masslike terms and a scalar potential,
beyond the standard gauge interactions; the possible gauge groups are
often severely restricted. For theories with a high
degree of supersymmetry, gaugings constitute the only known
supersymmetric deformations.   

In three spacetime dimensions the situation is special in two
respects. First of all, pure extended supergravity is
topological. Non-topological theories are constructed by coupling
supergravity to matter. In three dimensions the obvious matter
supermultiplets are scalar multiplets, so that the resulting
Lagrangians take the form of a nonlinear sigma model coupled to
supergravity. These theories have been constructed and classified
in~\cite{dWToNi93}. Supersymmetry leads to stringent conditions on the
target space, which can only be satisfied when the number of
supersymmetries is restricted to $N\leq16$, implying that there are at
most 32 supercharges. Beyond $N=4$ the target space has to be
homogeneous. There exists no theory with $N=7$ supersymmetry and beyond
$N=8$ there are only four possible
theories. They are $N=9,10,12$ and 16 supersymmetric, and their
target spaces are unique and equal to the symmetric spaces ${\rm
F}_{4(-20)}/{\rm SO}(9)$, ${\rm E}_{6(-14)}/({\rm SO}(10)\times {\rm
SO}(2))$, ${\rm E}_{7(-5)}/({\rm SO}(12)\times {\rm SO}(3))$ and ${\rm
E}_{8(8)}/{\rm SO}(16)$, respectively. 

Secondly, the gauging of these theories seems impossible at first
sight, because of the lack of vector gauge fields. However, one can
introduce a Chern-Simons term in three dimensions, which is
topological just as pure supergravity itself, and the corresponding
gauge fields can be coupled to the nonlinear sigma model by gauging a
subgroup of the target space isometries. Such gaugings have been
constructed in~\cite{NicSam00,NicSam01a,NicSam01b} for $N=16, 8$, and
in~\cite{DKSS99,AboSam01} for some abelian gauge groups for the case
of $N=2$. The gauging is defined by the gauge group embedding in the
isometry group, which in this case is defined in terms of a symmetric
embedding tensor. The latter defines the so-called $T$-tensors. The
viability of the gauging depends in a subtle manner on the properties
of these $T$-tensors and the gauged supergravity models have an
elegant mathematical structure (see~\cite{dWSaTr02} for the
corresponding analysis of the maximal supergravities in higher
dimensions). In this paper we exhibit this structure and derive the
precise conditions for having a consistent gauging. For $N>3$ these
conditions amount to the absence of $T$-tensor components transforming
in a particular irreducible $\SO{N}$ representation.
 
The gauged supergravities come with a scalar potential that allows for
supersymmetric anti-de Sitter groundstates. Therefore these theories can
be connected to two-dimensional superconformal theories that live on
the boundary of the anti-de Sitter space. The three-dimensional
setting may offer 
advantages when studying the adS/CFT correspondence, because
the supergravity theory is more amenable to nonperturbative studies,
while at the same time the large variety of two-dimensional
superconformal theories has been studied extensively in the
literature. The theories constructed in this paper include the
effective theories that arise in the compactification of
high-dimensional supergravities, as we will show below. These include
the compactifications on spheres~\cite{CvLuPo00,LuPoSe03},
compactifications with nontrivial
fluxes~\cite{GuVaWi99,HaaLou01,ACFH02,BeHaSa02}, as well as the
theories whose existence has been inferred from computing the
Kaluza-Klein spectra in the context of the adS-CFT
correspondence~\cite{BoPeSk98,DKSS98,dBoe98a,FuKeMi98,dBPaSk99}.  

The results of this paper constitute a complete classification of 
gauged supergravities in three dimensions which can be
regarded as an extension of the classification of ungauged
supergravities presented in \cite{dWToNi93}. In both cases the matter
supermultiplets comprise scalar and spinor fields. 
Because vector fields can always be converted to scalar fields by a
suitable duality transformation, the restriction to such scalar multiplets
does not seem relevant. However, the presence of gauge charges often poses
an obstacle for performing duality transformations. As it turns out,
no such obstacle arises in the three-dimensional context. Below we
will indicate how, by introducing compensating fields, every Yang-Mills
Lagrangian can be converted into a Lagrangian that belongs to the
class of  Lagrangians discussed in this paper. Therefore the
classification of the gauged supergravities presented here, is on a
par with the classification of the ungauged supergravities given
in~\cite{dWToNi93}, albeit that it is not quite possible to present an
exhaustive classification of all possible gauge groups. 

We briefly indicate how the conversion of three-dimensional Yang-Mills
Lagrangians can be done. In this conversion every gauge field is
replaced by two gauge fields and a new scalar field, which together
describe the same number of dynamic degrees of freedom as the original
gauge field. Our presentation is a further elaboration of the results
of~\cite{BeHaSa02,NicSam03a} and is completely general. Consider the
Lagrangian, in three spacetime dimensions, with Yang-Mills term
quadratic in the field strengths, and moment interactions proportional
to a gauge covariant operator $O^A_{\mu\nu}$,
\ba
\label{YM-L}
\CL = -\ft14\sqrt{g}
\left(F^A_{\mu\nu}(A)\pls O^A_{\mu\nu}(A,\Phi)\right)M_{AB}(\Phi)
\left(F^{B\mu\nu}(A)\pls O^{B\mu\nu}(A,\Phi)\right)  +
\CL^\prime(A, \Phi) \;.
 \ea
Here $A_\mu^A$ and $F_{\mu\nu}^A(A)$ denote the nonabelian gauge fields
and corresponding field strengths, labeled by indices $A,B,\ldots$,
and $\Phi$ generically denotes possible matter fields transforming
according to certain representations of the gauge group ${\rm G}_{\rm
YM}$. The structure constants of this group are denoted by
$f_{AB}{}^{\!C}$,  so that the field strengths read, 
\ba
F^A_{\mu\nu}(A) = \partial_\mu A_\nu^{A} - \partial_\nu A_\mu^{A}
-f_{BC}{}^{\!A}\, A_\mu^{B} A_\nu^{C} \;. 
\nn
\ea
The symmetric matrix $M_{AB}(\Phi)$ may depend on the matter fields
and transforms covariantly under ${\rm G}_{\rm YM}$. The last term,
$\CL^\prime(A,\Phi)$, in the Lagrangian is separately gauge invariant
and its dependence on the gauge fields is exclusively contained in
covariant derivatives of the matter fields or in topological mass
terms ({\it i.e.}\ Chern-Simons terms). The Bianchi identities and
vector field equations take the form,
\ba
\label{FB-eq}
D_\mu \tilde F^{A\,\mu}(A) = 0\;, \qquad
D^{\vphantom{B}}_{[\mu}\left(M_{AB}(\Phi)\, 
(\tilde F^B_{\nu]}(A)+\tilde O^B_{\nu]}(A,\Phi))\right) 
- J_{A\,\mu\nu}(A,\Phi) =0\;,
\ea
where we use the definitions
\ba
\tilde F_\mu^A(A) &=& \ft12 i \sqrt{g}\, \varepsilon_{\mu\nu\rho}
\,F^{A\,\nu\rho}(A) \;,\qquad 
J_{A\,\mu\nu}(A,\Phi) ~=~ \ft12 i \sqrt{g}\, \varepsilon_{\mu\nu\rho}
\,{\partial \CL^\prime(A,\Phi) \over\partial A_\rho^A}   \;,
\non[-.5ex]
\tilde O_\mu^A(A,\Phi) &=& \ft12 i \sqrt{g}\, \varepsilon_{\mu\nu\rho}
\,O^{A\,\nu\rho}(A,\Phi) \;. 
\nn
\ea
Usually the duality is effected by regarding the field strength as an
independent field on which the Bianchi identity is imposed by means of
a Lagrange multiplier. Because the Lagrangian \eqn{YM-L} depends
explicitly on both the field strengths and on the gauge fields, we
proceed differently and write the field strength in terms of new
vector fields $B_{A\,\mu}$ and the derivative of compensating scalar
fields $\phi_A$, all transforming in the adjoint representation of the
gauge group. The explicit expression,
\ba
\label{F-B}
\tilde F_\mu^A(A)+\tilde O_\mu^A(A) &=& 
M^{AB}(B_{B\,\mu} - D_\mu \phi_B)\;,
\ea
where $M_{AC}\,M^{CB}= \delta^B_A$, should be regarded as a field
equation that follows from the new Lagrangian that we are about to
present. The structure of \eqn{F-B} implies that we are dealing with 
additional gauge transformations as its right-hand side is invariant
under the combined transformations, 
\ba
\delta B_{A\,\mu} = D_\mu \Lambda_A \,,\qquad \delta \phi_A =
\Lambda_A
\;,
\ea
under which all other fields remain invariant. The corresponding
abelian gauge group, $\cal T$, has nilpotent generators transforming
in the adjoint representation of ${\rm G}_{\rm YM}$. Obviously, the
$\phi_A$ act as compensating fields with respect to $\cal
T$. The combined gauge group is now a semidirect product of ${\rm
G}_{\rm YM}$ and $\cal T$ and its dimension is twice the dimension of 
the original gauge group 
${\rm G}_{\rm YM}$. The covariant field strengths belonging to the new
gauge group are 
$F^A_{\mu\nu}{}(A)$ and $F_{A\,\mu\nu}(B,A)=
2\,D_{[\mu}B_{A\,\nu]}$, and transform under $\cal T$ 
according to $\delta F^A_{\mu\nu}{}=0$ and $\delta F_{A\,\mu\nu}=
- \Lambda_C \, f_{AB}{}^{\!C}\,F^B_{\mu\nu}{}$. The fully gauge
covariant derivative of $\phi_A$ equals 
\ba
{\hat D}_\mu \phi_A &\equiv&
D_\mu\phi_A  - B_{A\,\mu} ~=~  \partial_\mu\phi_A -
f_{AB}{}^{\!C} \,A^B_\mu{}\,\phi_C  - B_{A\,\mu}\;,
\ea
and is invariant under ${\cal T}$ transformations. 

The field equations corresponding to the new Lagrangian,
\ba
\CL &=& -\ft12 \sqrt{g}\,\hat{D}_\mu \phi_A
\,M^{AB}(\Phi)\hat{D}^\mu \phi_B 
+\ft12 i \varepsilon^{\mu\nu\rho} \,
(F^A_{\mu\nu}B_{A\,\rho} -
O^{A}_{\mu\nu}\, \hat{D}_\rho \phi_A )
+ \CL^\prime(A, \Phi) \;, 
 \ea
lead to \eqn{F-B} and \eqn{FB-eq}, as well as to the same field equations
as before for the matter fields $\Phi$. Observe that the Lagrangian is
fully gauge invariant up to a total derivative. Hence, the Yang-Mills
Lagrangian has now been converted to a Chern-Simons Lagrangian, with a
different gauge group and a different scalar field content. To obtain
the original Lagrangian \eqn{YM-L}, one simply imposes the gauge
$\phi_A=0$ and integrates out the fields $B_{A\,\mu}$. 

This paper is organized as follows. In section~2 we briefly summarize
the results of~\cite{dWToNi93} for the ungauged theories. Subsequently
we analyze the possible invariances of the Lagrangian possibly related
to target space isometries.  Then, in section~3, we discuss the
gauging of possible subgroups of the isometry group. We determine the
potential and the masslike terms in the general case and derive the
extra conditions that must be satisfied in order to preserve
supersymmetry. This is the central result of this paper. In section~4
we analyze these restrictions in detail for $N\leq4$. For $N>4$ the
target spaces are homogeneous, and their consistent gaugings are
discussed in section~5. We present some concluding remarks in
section~6. Some technical details are relegated to two appendices.

\mathon
\section{Nonlinear sigma models coupled to supergravity}
\mathoff
In this section we summarize and elaborate on the construction of
three-dimensional nonlinear sigma models coupled to supergravity. For
the derivation and conventions we refer to \cite{dWToNi93}. The fields
of the nonlinear sigma model are scalar fields~$\phi^i$ and spinor
fields~$\chi^i$, with $i=1, \ldots, d$; the supergravity fields are
the dreibein $e_\mu{}^a$, the spin-connection field~$\omega^{ab}_\mu$
and $N$ gravitini fields~$\psi^I_\mu$ with $I=1,\ldots,N$. The
gravitini transform under the R-symmetry group ${\rm SO}(N)$, which is
not necessarily a symmetry group of the Lagrangian. The scalar fields
parametrize a target space endowed with a Riemannian metric
$g_{ij}(\phi)$.

\subsection{Target-space geometry}

Pure supergravity is topological in three dimensions and exists for an
arbitrary number $N$ of supercharges and corresponding gravitini
\cite{AchTow86}. Its coupling to a nonlinear sigma model requires the
existence of $N-1$ almost complex structures $f^{Pi}{}_j(\phi)$,
labeled by $P= 2, \ldots,N$, which are hermitean,
\ba
g_{ij}\,f^{Pj}{}_k +g_{kj}\,f^{Pj}{}_i  = 0\,, 
\ea
and generate a Clifford algebra, 
\ba
f^{Pi}{}_k \,f^{Qk}{}_j +  f^{Qi}{}_k\, f^{Pk}{}_j = -2\,
\delta^{PQ}\,\delta^i_j \,.
\ea
{}From the $f^P$ one constructs
$\ft12N(N\mis1)$ tensors $f^{IJ}_{ij}=-f^{JI}_{ij}= -f^{IJ}_{ji}$ that
can act as generators for the group ${\rm SO}(N)$,
\ba
f^{PQ}= f^{[P}\,f^{Q]} \,,\qquad f^{1P}= -f^{P1} = f^P \,,
\ea
where, here and  henceforth, $I,J=1,\ldots,N$. The tensors $f^{IJ}$
satisfy (in obvious matrix notation),
\ba
f^{IJ}\,f^{KL} &=& f^{[IJ}\,f^{KL]} - 4\,\delta^{[I[K}\,f^{L]J]} -
2\,\delta^{I[K}\,\delta^{L]J} \,{\bf 1} \,, \nonumber \\[1ex]
f^{IJ\,ij}\,f^{KL}{}_{\!ij}  &=& 2d \,\delta^{I[K}\,\delta^{L]J} - 
\delta_{N,4}\, \varepsilon^{IJKL} \, {\rm Tr}(J)\,.
\label{f-properties}
\ea
Only for $N=4$, the tensor $J^i{}_j$ is relevant; it is defined by $J=
\ft16\,\varepsilon_{PQR}\,f^P\,f^Q\,f^R$, so that 
\be
f^P\,f^Q = -\delta^{PQ} \,{\bf 1} - \varepsilon^{PQR}\,J\, f^R \,. 
\ee
Furthermore, $J$ satisfies,
\ba
J\,f^{P}= f^{P}\,J \,,\qquad J^2 = {\bf 1} \,, \qquad J_{ij} =
J_{ji}\,, \qquad J=\ft1{24} \varepsilon^{IJKL} f^{IJ} f^{KL} 
\;, 
\ea
and has eigenvalues equal to $\pm 1$. It turns out
that $J$ is also covariantly constant, which implies that the target
space is locally the product of two separate Riemannian spaces of
dimension $d_\pm$, where $d_++d_-=d$ and $d_\pm$ are both multiples of
4. These two spaces correspond to inequivalent $N=4$
supermultiplets. Hence the case $N=4$ is rather special, and the
last identity \eqn{f-properties} can be written as 
\be
f^{IJ\,ij}\,f^{KL}{}_{\!ij}  = 4 \left(d_+\,\bbP_+^{IJ,KL} +
d_-\,\bbP_-^{IJ,KL}\right) \;,
\ee
with projectors,
\ba
\bbP_\pm^{IJ,KL} &=& \ft12 \delta^{I[K}\,\delta^{L]J} \mp\ft14 
\varepsilon^{IJKL} \;.
\label{P4}
\ea

For rigidly supersymmetric nonlinear sigma models, the number of
supersymmetries is equal to $N=1,2$ or 4, and the Lagrangians are
manifestly invariant under ${\rm SO}(N)$ R-symmetry transformations
acting exclusively on the fermion fields through multiplication with
the complex structures. The case $N=3$ is not distinct from $N=4$,
because the existence of two complex structures necessarily implies
the existence of a third one. In case of $N=4$, the Lagrangian
is a sum of two separate Lagrangians corresponding to the
$d_\pm$-dimensional target spaces. For $N=3,4$ the target spaces are
hyperk\"ahler.  

When coupling to supergravity the Lagrangian and supersymmetry
transformations depend on ${\rm SO}(N)$ target-space connections
denoted by $Q_i^{IJ}(\phi)$. These connections are nontrivial in view
of 
\ba
R^{IJ}_{ij}(Q)\equiv \partial^{~}_i Q_j^{IJ} - \partial_j Q_i^{IJ} + 2
Q_i^{K[I}Q_j^{J]K} &=& \ft12 f^{IJ}_{ij} \;.
\label{fQ}
\ea
For local supersymmetry $N$ can take the values $N=1,\ldots,6$ and
$8,9,10,12$ or $16$. The situation regarding ${\rm SO}(N)$ symmetry is
more subtle in this case, as we shall discuss in due course. The $N=3$
theory is no longer equivalent to an $N=4$ theory, as it has
only three
gravitini. In view of the three almost complex structures, the target
space is a quaternionic space. For $N=4$ the target space decomposes
locally into a product of two quaternionic spaces of dimension
$d_\pm$. The $f^{IJ}$ are covariantly constant, both with respect to
the Christoffel and the ${\rm SO}(N)$ connections, $\Gamma_{ij}{}^k$
and $Q_i^{IJ}$, respectively,
\ba
\label{Q-curv}
D^{~}_i(\Gamma,Q)\,f^{IJ}_{jk}\equiv \partial_i f^{IJ}_{jk}
- 2\, \Gamma^{~}_{i[k}{}^l\, f^{IJ}_{j]l}  +2\, 
Q^{K[I}_i\, f^{J]K}_{jk} =0\,.
\ea
For $N>2$ we are thus dealing with almost complex, rather than with
complex, structures. This implies an integrability condition for the
target-space Riemann tensor $R_{ijkl}$, 
\ba
\label{integrability}
R^{~}_{ijmk}\,f^{IJ\,m}{}_{\!l} -R_{ijml}\,
f^{IJ\,m}{}_{\!k} = -f^{K[I}_{ij}  \,f^{J]K}_{kl} \,,
\ea
where we made use of \eqn{fQ}. Contracting \eqn{integrability} with
$f^{MNkl}$ gives, for general $N>2$,  
\ba
\label{holonomy}
R_{ijkl}\,f^{IJ\,kl} = \ft14d\,f^{IJ}_{ij} \,, 
\ea
so that the target space has nontrivial ${\rm SO}(N)$ holonomy, 
while contracting \eqn{integrability} with~$g^{jl}$, using the
cyclicity of the Riemann tensor and the above result, yields (for
$N>2$) 
\ba
\label{Einstein}
R_{ij} \equiv R_{ikjl}\,g^{kl} = c\, g_{ij} \,,
\ea
with $c= N -2 +\ft18 d > 0$.
Hence the target space must be an Einstein space.\footnote{
  For $N=3$ this is in accord with the fact that quaternionic spaces
  of $d>4$ are always Einstein \cite{Alek68}. In
  the case at hand, the result also holds true for a $d=4$ 
  target space. For $N=4$ the equations \eqn{holonomy} and
  \eqn{Einstein} read 
\ba
R_{ijkl}\,f^{IJ\,kl} &=& \ft12\left(d_+\,\bbP_+{}^{IJ,KL} +
d_-\,\bbP_-{}^{IJ,KL}\right) f_{ij}^{KL}\;, \nonumber\\
R_{ij} &=& (2 +\ft18 d)\, g_{ij} + \ft18 (d_+-d_-) J_{ij} \;,
\nn
\ea
and we have a product space of two quaternionic manifolds, which are
both Einstein. For $N=2$ the target space is K\"ahler and $f^{IJ}$ is
a complex structure. The ${\rm SO}(2)$ holonomy is undetermined. } 

Following \cite{dWToNi93} we introduce a complete set of linearly
independent, antisymmetric, tensors $h^\alpha_{ij}$, labeled by
indices $\alpha$, 
that commute with the complex structures, {\it i.e.}, 
\be
h^\alpha_{ik}\,f^{IJ\,k}{}_j -h^\alpha_{ik}\,f^{IJ\,k}{}_j =0\,.
\label{h-tensors}
\ee
For $N=2$, there is only one tensor $f^{IJ}$ which commutes with
itself, so that this decomposition is not meaningful. For $N>2$ we
must have $h^\alpha_{ij}\,f^{IJ\,ij}=0$.  The tensors $h^{\alpha
i}{}_{j}$ generate a subgroup ${\rm H}^\prime \subset {\rm SO}(d)$
that commutes with the group ${\rm SO}(N)$ generated by the
tensors~$f^{IJ}$.  They can be normalized according to
$h^{\alpha}_{ij}h^{\beta \,ij}\propto \delta^{\alpha\beta}$ and are
covariantly constant with respect to the Christoffel connection and a
new connection $\Omega_i^{\alpha\beta}$,
\be
D_i(\Gamma) h^\alpha_{jk} - \Omega_i^{\alpha}{}_\beta\,h^\beta_{jk}=0\,.
\ee
The Riemann tensor can be written as ($N>2$)
\ba
\label{curvature-constraint}
R_{ijkl} &=& 
\frac18\,\Big( f^{IJ}_{ij} \,f^{IJ}_{kl} + C_{\alpha\beta}
\,h^\alpha_{ij}\,h^\beta_{kl}\Big) \;, 
\ea
where $C_{\alpha\beta}(\phi)$ is a symmetric tensor. This result
implies that the holonomy group is contained in ${\rm
SO}(N)\times {\rm H}^\prime\subset {\rm SO}(d)$ which must act
irreducibly on the target space. For $N=4$ this result is modified
because of the product structure. Observe that the Bianchi identities
on the curvature tensor are not manifest for the expression on the
right-hand side of \eqn{curvature-constraint}, something that plays an
important role in the analysis of \cite{dWToNi93}.The ${\rm H}^\prime$
curvatures can now be shown to take the form,
\ba
2(\partial_{[i}\Omega_{j]}{}^\alpha{}_\beta
-\Omega_{[i}{}^\alpha{}_\gamma \,\Omega_{j]}{}^\gamma{}_\beta) = \ft18
f^{\alpha \gamma}{}_\beta \,C_{\gamma\delta} \,h^\delta_{ij}\;,
\ea
where we have made use of the structure constants of the group ${\rm
H}^\prime$ defined by,
\ba
h^\alpha\,h^\beta -h^\beta\,h^\alpha = f^{\alpha\beta}{}_\gamma\, h^\gamma\;.
\ea
The above result shows that the connections $\Omega_i{}^{\alpha\beta}$
can be restricted to the form $\Omega_{i}{}^{\alpha\beta}\propto
f^{\alpha\beta}{}_\gamma\, Q_i{}^\gamma$. 
\subsection{Lagrangian and invariances}
\label{lagrangian}
Let us now turn to the Lagrangian and supersymmetry transformations.
In the following it is convenient to adopt an ${\rm SO}(N)$ covariant
notation which allows to select the $N\mis1$ almost complex structures
from the $f^{IJ}$ tensors by specifying some arbitrary unit $N$-vector
$\alpha_I$ and identifying the complex structures with $\alpha_J
f^{JI}$. By extending the fermion fields $\chi^i$ to an overcomplete
set $\chi^{iI}$, defined by
\ba
\chi^{iI} = \left(\chi^i, f^{Pi}{}_j\,\chi^j \right) \;,
\ea
we can write the Lagrangian and transformation rules in a way that
does no longer depend explicitly on the almost-complex structures. The
fact that we have only $d$ fermion fields, rather than $dN$, can be
expressed by the ${\rm SO}(N)$ covariant constraint, 
\ba
\chi^{iI} &=& \bbP_J^I{}_j^i\,  \chi^{jJ} ~\equiv~
{1\over N}\left(\Gd^{IJ}\Gd^{i}_j - f^{IJ\,i}{}_j\right) \chi^{jJ}
\;.
\label{projchi}
\ea
The trace of this projector equals $\bbP^I_I{}^i_i=d$, which confirms
that the total number of fermion fields is not altered. We should
stress here, that the introduction of $\chi^{iI}$ is a purely
notational exercise and we do not aim at implementing the constraint
\Ref{projchi} at the Lagrangian level. At every step in the
computation one may change back to the original notation by choosing
$\chi^i = \alpha_I\chi^{iI}$. The covariant notation does not imply
that the theory is ${\rm SO}(N)$ invariant, but the covariant setting
allows us to treat the $N$ supersymmetries and the corresponding
gravitini on equal footing and it facilitates the various derivations
in later sections.

The supersymmetry transformations read 
\begin{eqnarray}
\Gd e_\mu{}^a &=& \ft12\,\Be{}^I\Gg^a\,\psi^I_\mu 
\;,\non
\Gd \psi^I_\mu &=& D_\mu \epsilon^I   - \ft18 g_{ij} \,\Bchi{}^{iI}
\gamma^\nu \chi^{jJ}\, \gamma_{\mu\nu} \,\epsilon^J - 
\Gd \phi^i \,Q_i^{IJ} \psi^J_\mu \;,\non
\Gd \phi^i &=& \ft12\,\Be^I\,\chi^{iI} 
\;,\non
\Gd \chi^{iI} &=& \ft12 \left(\delta^{IJ}{\bf 1} \mis f^{IJ}
\right)^i{}_{\!j}\; 
\FMslash{\widehat\partial} \phi^j \, \epsilon^J -
\Gd \phi^j \left(
\Gamma^i_{jk}\,\chi^{kI} 
+ Q_j^{IJ}\chi^{iJ} \right)
\;,
\label{susytraf}
\end{eqnarray}
where the supercovariant derivative $\widehat\partial_\mu \phi^i$ and
the covariant derivative $D_\mu(\omega,Q)\,\epsilon^I$ are defined by
\ba
  \widehat \partial_\mu \phi^i &=& \partial_\mu \phi^i -\ft12
\Bpsi{}^I_\mu \chi^{iI} \;,\nonumber\\
D_\mu \epsilon^I &=& \left( \partial_\mu +\ft12 \omega^a_\mu
\,\gamma_a \right )\epsilon^I +\partial_\mu \phi^i\, Q_i^{IJ}
\,\epsilon^J\;.  
\ea
Observe that the terms proportional to $\delta \phi$  in 
$\delta\chi^{iI}$ do not satisfy the same constraint
\eqn{projchi} as $\chi^{iI}$ itself, because the projection operator 
$\bbP_J^I{}_j^i$ itself transforms under supersymmetry. As in
\cite{dWToNi93},  we use the Pauli-K\"all\'en metric with hermitean
gamma matrices $\gamma^a$, satisfying  
$\Gg_a\Gg_b = \delta_{ab} +\I \Geps_{abc} \Gg^c$.  

Let us now turn to the Lagrangian, which reads
\ba
\CL_0 &=& -\ft12\,\I\,\varepsilon^{\mu\nu\rho}\left(
e_\mu{}^a\,R_{\nu\rho a} +
\Bpsi{}^I_\mu D_\nu\psi^I_\rho  \right)
-\ft12 e\,g_{ij} \left( g^{\mu\nu}\, \partial_\mu \phi^i\,
\partial_\nu \phi^j + N^{-1} \Bchi{}^{iI} \FMslash{D}
\chi^{jI} \right)\non[1ex]
&&{}+ \ft14 e\, g_{ij}\,  \Bchi{}^{iI}\Gg^\mu\Gg^\nu
\psi_\mu^I\,(\dd_\nu\phi^j+\widehat \partial_\nu \phi^j)  
-\ft1{24} e \,N^{-2} R_{ijkl} \; \Bchi{}^{iI} \gamma_a
\chi^{jI} \, \Bchi{}^{kJ} \gamma^a \chi^{lJ} \non[1ex] 
&&{} +  \ft1{48} e\,N^{-2} \left( 3\,( g_{ij}\, \Bchi{}^{iI}
\chi^{jI})^2 - 2(N-2)\, ( g_{ij}\, \Bchi{}^{iI}\gamma^a 
\chi^{jJ})^2  \right)
\;,
\label{L}
\end{eqnarray}
Here we used the covariant derivatives
\begin{eqnarray}
  D_\mu \psi^I_\nu &=& \left( \partial_\mu +\ft12
\omega^a_\mu \,\gamma_a \right ) \psi^I_\nu + 
\partial_\mu \phi^i \, Q_i^{IJ} \psi^J_\nu\;, \non[.5ex]
  D_\mu \chi^{iI} &=&  
\left( \partial_\mu +\ft12 \omega^a_\mu\, \gamma_a \right)\chi^{iI} 
+\partial_\mu \phi^j \left( \Gamma^i_{jk}\, \chi^{kI} +
Q_j^{IJ} \chi^{iJ} \right)  \;. 
\end{eqnarray}
We emphasize that the above results coincide with the results of 
\cite{dWToNi93}, written in a different form. The conversion makes use
of \eqn{integrability}. The Lagrangian and transformation rules are
consistent with target-space diffeomorphisms and field-dependent 
${\rm SO}(N)$ R-symmetry rotations, acting on $\psi^I_\mu$,
$\chi^{iI}$  and $Q_i^{IJ}$ according to 
\ba
\delta\psi^I_\mu = \Lambda^{IJ}(\phi)\, \psi^J_\mu\,,\qquad 
\delta\chi^{iI} = \Lambda^{IJ}(\phi)\, \chi^{iJ}\,,\qquad 
\delta Q_i^{IJ} = - D_i \Lambda^{IJ}(\phi)\,.
\label{so-1}
\ea
Combining the last result with \eqn{fQ}, one concludes that 
the $f^{IJ}$ should also be rotated,
\ba
\delta f^{IJ} = 2\,\Lambda^{K[I}(\phi) \,f^{J]K}\,.
\label{so-2}
\ea 
The target-space diffeomorphisms and field-dependent 
${\rm SO}(N)$ R-symmetry rotations correspond to reparametrizations
within certain equivalence classes, but do not, in general, constitute
an invariance.

In the remainder of this section we discuss the invariances of these
models, other than supersymmetry, spacetime diffeomorphisms and local
Lorentz transformations.  The target space may have isometries,
generated by Killing vector fields $X^i(\phi)$. Some of them can be
extended to invariances of the full Lagrangian, possibly after
including a field-dependent ${\rm SO}(N)$ transformation according to
\eqn{so-1} and \eqn{so-2}. Hence, we combine an isometry characterized
by a Killing vector field $X^{i}$ with a special ${\rm SO}(N)$
transformation whose parameters depend on $X^{i}(\phi)$ and on the
scalar fields.  Denoting the infinitesimal ${\rm SO}(N)$
transformations by ${\cal S}^{IJ}(X,\phi)$, we require invariance of
the target-space metric, the ${\rm SO}(N)$ connections and the almost
complex structures, up to a uniform ${\rm SO}(N)$ rotation, {\it
i.e.},
\ba
\CL_{X}\, g_{ij} &=&  0 \;, \non
\CL_{X} Q_i^{IJ} + D_i {\cal S}^{IJ}(\phi,X)&=& 0 \;,\non
\CL_{X} f_{ij}^{IJ} - 2\, {\cal S}^{K[I}(\phi,X)\,f^{J]K}_{ij} 
&=& 0 \;. 
\label{isomf}
\ea
The Lagrangian \Ref{L} is then invariant under the combined
transformations, 
\ba
\label{iso-trans1}
\Gd\phi^i      = X^i(\phi) \;,\quad
\Gd \psi^I_\mu = {\cal S}^{IJ}(\phi,X) \,\psi_\mu^J \;,\quad
\Gd \chi^{iI}  = \chi^{jI}\partial_jX^i +
{\cal S}^{IJ}(\phi,X) \, \chi^{iJ} \;.
\ea
The fermion transformations can be rewritten covariantly, 
\ba
\label{iso-trans2}
\Gd \psi^I_\mu &=& {\cal V}^{IJ}(\phi,X) \,\psi_\mu^J - \Gd\phi^i\,
Q_i^{IJ} \psi^J_\mu \;,\non
\Gd \chi^{iI}  &=& D_jX^i \,\chi^{jI}+
{\cal V}^{IJ}(\phi,X) \, \chi^{iJ} -
\Gd \phi^j \left(\Gamma^i_{jk}\,\chi^{kI} 
+ Q_j^{IJ}\chi^{iJ} \right) \;,
\ea
where $\CV^{IJ}(\phi,X)\equiv X^j Q_j^{IJ}(\phi) +
{\cal S}^{IJ}(\phi,X)$. 
The significance of this result will become apparent in a
sequel. Using \eqn{fQ} and \eqn{Q-curv}, one verifies
that the second and third equation of \Ref{isomf} can be written as, 
\ba
D_i \CV^{IJ}(\phi,X) &=& \ft12 f^{IJ}_{ij}(\phi)\, X^j(\phi) \;,\non
f^{IJk}{}_{[i}(\phi) \;D_{j]}X_k(\phi) &=& f_{ij}^{K[I}(\phi)
\;\CV^{J]K}(\phi,X) \;. 
\label{DXS}
\ea
The first equation in~\Ref{DXS} shows that $\CV^{IJ}(\phi,X)$ can be
regarded as as the moment map associated with the isometry $X^i$. The
second equation is just the integrability condition of
the first equation, so it is not independent. After contraction with
$f^{MN\,ij}$, it leads to
\ba
f^{IJ\,ij}\,D_iX_j  &=& \left\{
\begin{array}{ll}
\ft12 d\,\CV^{IJ} \;, & \mbox{for $N\not= 2,4$} \\[1ex]
(d_+\,\bbP_+^{IJ,KL} + d_-\,\bbP_-^{IJ,KL})
\CV^{KL}\;,  & \mbox{for $N=4$}
\end{array}
\right.
\label{S}
\ea
{}From combining the above equations it follows that $\Delta \CV^{IJ} =
\ft14\, d\, \CV^{IJ}$, where $\Delta$ denotes the ${\rm SO}(N)$
covariant Laplacian. This result applies to $N>2$, with obvious
modifications for $N=4$.  The above analysis shows that there are no
restrictions for $N>2$ to extend an isometry to a symmetry of the
Lagrangian. For $N=2$ this is different, as the isometry should be
holomorphic, {\it i.e.}, it should leave the complex structure
invariant. In this case $\CV^{IJ}$ is determined by \Ref{DXS} up to an
integration constant. This constant is related to an invariance under
constant $\SO2$ transformations of the fermions.

For $N=4$ we note that the complex structures
$\bbP_\pm^{IJ,KL}\,f^{KL}$ live in the two separate quaternionic
subspaces. The same is true for $\bbP_\pm^{IJ,KL}\,\CV^{KL}$, which
according to \eqn{DXS} depends only on the corresponding subspace
coordinates. Note, however, that when one of the subspaces is trivial,
say when $d_-=0$, then $\bbP_-^{IJ,KL}\,\CV^{KL}$ corresponds to a
triplet of arbitrary constants. This is a consequence of the fact that
the model in this case has a rigid $\SO3$ invariance which acts
exclusively on the fermions.

The supersymmetry transformations do not commute with the isometries,
as one can verify most easily on the fields $\phi^i$, where one derives
\be
\label{QG-comm}
{[}\delta_{\rm Q}(\epsilon), \delta_{\rm G}(X)] = \delta_{\rm
Q}(\epsilon^\prime)\;,\qquad
\mbox{with}\quad 
(\epsilon^I)^\prime = {\cal S}^{IJ}(\phi,X) \,\epsilon^J\;. 
\ee

The isometries that can be extended to an invariance of the
Lagrangian, generate an algebra ${\mathfrak g}$. Denoting $\{X^\cM\}$
as a basis of generators, we have 
\ba 
X^{\cM i}\,\dd_i X^\cN - X^{\cN i}\,\dd_i X^\cM ~=~
f^{\cM\cN}{}_{\!\!\cK}\,X^\cK \;,
\label{fABC}
\ea
with structure constants $f^{\cM\cN}{}_{\!\!\cK}$. 
Closure of the algebra implies that the corresponding induced $\SO{N}$
rotations, ${\cal S}^{\cM\,IJ}\equiv{\cal S}^{IJ}(\phi, X^{\cM})$,
satisfy,  
\begin{equation}
\label{SABC}
  \left[{\cal S}^{\cM}, {\cal S}^{\cN}\right]^{IJ} =
- f^{{\cM}{\cN}}{}_{\!\!\cK} \,{\cal S}^{\cK\,IJ} +
\left(X^{{\cM}i} \,\partial_i {\cal S}^{\cN\,IJ} -
X^{{\cN}i} \,\partial_i {\cal S}^{\cM\,IJ}\right)\;.
\end{equation}
Using \Ref{fABC} and the second equation \eqn{isomf}, this can be
rewritten as 
\ba
\left[ \CV^\cM  ,\, \CV^\cN \right]^{IJ} = - 
f^{{\cM}{\cN}}{}_{\!\!\cK} \,\CV^{\cK\,IJ} + 
\ft12\,f^{IJ}_{ij}\,X^{\cM i}X^{\cN j}   \;,
\label{ABCSON}
\ea
with $\CV^{\cM IJ}\equiv\CV^{IJ}(\phi, X^{\cM})$. In the case of $N=2$
and of $N=4$ with $d_+d_-=0$, the R-symmetry, which acts only on the
fermions, is realized as a separate invariance. Obviously, R-symmetry
commutes with the isometry group; for $N=2$ the R-symmetry may define
a central extension of the isometry group, while for $N=4$ with
$d_+d_-=0$, the invariance group takes the form of a direct product of 
the isometry 
group with an ${\rm SO}(3)$ factor of the R-symmetry group. This
implies that there are generators for which the Killing vector
$X^{\cM\,i}$ vanishes and ${\cal S}^{\cM IJ}$ is constant. We return
to this in our discussion of the specific cases in section~4. 

We now note that the second equation of \eqn{DXS} implies that $D_i
X_j - \ft14 \,f^{MN}_{ij}\,\CV^{MN}$ commutes with the almost complex
structures, so that it can be decomposed in terms of the antisymmetric
tensors $h^\alpha_{ij}$ that were introduced in \eqn{h-tensors},
\ba
D_i X^\cM_j - \ft14
f^{IJ}_{ij}\,\CV^{\cM\,IJ} &\equiv& h^\alpha_{ij}\,\CV^\cM{}_\alpha
\;.
\label{V2}
\ea
This result and the results given in the remainder of this section
apply only to $N>2$. 
Using the general result for Killing vectors, $D_iD_jX_k = R_{jkil}
\,X^l$, we can evaluate the derivative of
$\CV^{\cM}{}_\alpha$. Introducing furthermore the notation $\CV^{\cM
\,i}\equiv X^{\cM\,i}$, we establish the following system of linear
differential equations,
\ba
D_i \CV^{\cM\,IJ} &=& \ft12\,f^{IJ}_{ij}\,\CV^{\cM\, j} \;,
\non[.5ex]
D_i  \CV^\cM{}_{j} &=& \ft14\,f^{IJ}_{ij}\,\CV^{\cM\,IJ} 
+ h^\alpha_{ij}\,\CV^\cM{}_\alpha \;,
\non[.5ex]
D_i  \CV^\cM{}_{\alpha} &=&
\ft18\,C_{\alpha\beta}\,h^\beta_{ij}\,\CV^{\cM\,j}  \;,
\label{DV}
\ea
where the covariant derivative contains the Christoffel connection as
well as the ${\rm SO}(N)\times {\rm H}^\prime$ connections. One can
prove that $C_{\alpha\beta}$ is covariant under the isometry, {\it
i.e.}, $\CV^{\cM\,i} D_iC_{\alpha\beta}=2\, \CV^\cM{}_\gamma \,
f^{\delta\gamma}{}_{(\alpha}\,C_{\beta) \gamma}$. Other than that the
integrability of the above equations is guaranteed by previous
results. 
Furthermore, by substituting the second identity of \eqn{DV} into
\eqn{fABC}, and by taking the derivative of \eqn{fABC} and exploiting
previous identities, we derive the following two equations,
\ba
\label{ABCH}
f^{\cM\cN}{}_\cK\, \CV^{\cK}{}_i &=& \ft 14 f^{IJ}_{ij} 
(\CV^{\cM\,IJ} \,\CV^{\cN\,j}- \CV^{\cN\,IJ} \,\CV^{\cM\,j})  
+h^{\alpha}_{ij} 
(\CV^{\cM}{}_\alpha \,\CV^{\cN\,j}- \CV^{\cN}{}_\alpha
\,\CV^{\cM\,j})   \;,\non[.5ex]
f^{\cM\cN}{}_\cK\, \CV^\cK {}_\alpha &=&  f^{\beta\gamma}{}_\alpha \,
\CV^\cM{}_\beta\, \CV^\cN{}_\gamma + \ft18 C_{\alpha\beta} \, h^\beta_{ij}
\,\CV^{\cM\,i} \,\,\CV^{\cN\,j} \;.
\ea
 
The quantities $\CV^{\cM\,IJ}$, $\CV^{\cM\,i}$ and $\CV^{\cM}{}_\alpha$
transform according to the adjoint representation of the invariance
group, up to field-dependent ${\rm SO}(N)\times {\rm H}^\prime$
transformations, as is shown by (note that for $\CV^{\cM\,i}$ this
already follows from \eqn{fABC}),
\ba
\label{cov-CV}
\CV^{\cN\,i}D_i\, \CV^{\cM\,IJ} &=& - f^{\cM\cN}{}_\cK\, \CV^{\cK\,IJ} +
[\CV^\cN,\CV^\cM]^{IJ}\, \non
\CV^{\cN\,i}D_i\, \CV^{\cM}{}_\alpha &=& - f^{\cM\cN}{}_\cK\,
\CV^{\cK}{}_\alpha  + f^{\beta\gamma}{}_\alpha \,\CV^\cN{}_\gamma\,
\CV^\cM{}_\beta \,.
\ea

We close this section with a few observations regarding the structure
of the last equations \eqn{ABCH} and \eqn{cov-CV}. Let us 
first note that the following operators   
\ba
{\cal D}^\cM = \delta^i_j\,\CV^{\cM\,k}\,D_k  + \ft14 f^{IJ\,i}{}_j
\CV^{\cM\,IJ} + h^{\alpha\,i}{}_j\, \CV^\cM{}_\alpha\;,
\ea
acting in the space of target-space tensors, provide a realization of
the Lie algebra $\mathfrak{g}$ associated with the invariance group,
according to
\ba
{[}{\cal D}^\cM ,{\cal D}^\cN] = f^{\cM\cN}{}_\cK\;{\cal D}^\cK\;.
\ea

The equations \eqn{ABCH} can also be encoded in the following
algebraic structure. Define the algebra  
$\mathfrak{a}\equiv\{t^{\cA}\}\equiv \{t^{IJ}, t^{\Ga}, t^{i}\}$,
as an extension of $\mathfrak{so}(N)\oplus \mathfrak{h'}$ with
commutation relations, 
\ba
\left[t^{IJ},t^{KL}\right] &=& -4\, \delta\oversym{^{I[K}\,t^{L]J}\,}
\;,
\qquad
\left[t^{\Ga},t^{\Gb}\right] ~=~ f^{\Ga\Gb}{}_{\!\Gg}\,t^\Gg \;,
\qquad
\left[t^{IJ},t^{i}\right] ~=~ \ft12\,f^{IJ}{}_{\!j}{}^i\,t^j\;,
\non[1ex]
\left[t^{\Ga},t^{i}\right] &=& h^{\Ga}{}_{\!j}{}^i\,t^j\;,
\qquad\qquad\quad
\left[t^{i},t^{j}\right] ~=~ 
\ft14\,f_{IJ}^{ij}\,t^{IJ} + \ft18\,C_{\Ga\Gb}h^{\Gb\,ij}\,t^\Ga
\;.
\label{algebra}
\ea
Unless $C_{\alpha\beta}$ is an ${\rm H}^\prime$-invariant tensor, this
algebra will be nonassociative (or may alternatively be realized as a
soft associative algebra upon imposing $[t^\alpha,C_{\beta\gamma}] = -2
f^{\alpha\delta}{}_{\!(\beta}\,C_{\gamma)\delta}$). 
Equations \eqn{ABCSON} and \eqn{ABCH} then imply that the map,
\ba
\CV:~ \mathfrak{g} &\rightarrow& \mathfrak{a}\,,\qquad
\CV(X^\cM)~:=~ \CV^{\cM}{}_{\!\cA}\,t^{\cA} ~=~
\ft12\CV^{\cM}{}_{\!IJ}\,t^{IJ} +
\CV^{\cM}{}_{\!\Ga}\,t^{\Ga} + \CV^{\cM}{}_{\!i}\,t^{i} \;, 
\label{Vmap}
\ea 
defines a Lie algebra homomorphism, {\it i.e.}\
$\CV([X^\cM,X^\cN])=[\CV(X^\cM),\CV(X^\cN)]$. 
In particular, the image of $\mathfrak{g}$ under $\CV$ is an associative
subalgebra of $\mathfrak{a}$. Furthermore, \eqn{DV} 
takes the simple form,
\ba
D_i \CV(X^\cM) &=& \left[ g_{ij}\,t^j , \CV(X^\cM) \right] \;.
\ea
When the inverse $C^{\alpha\beta}$ exists, one can prove that 
\ba
\ft12 \CV^{\cM\,IJ}\,\CV^{\cN\,IJ} + \CV^\cM{}_i\,\CV^{\cN\,i} -8
\,C^{\alpha\beta} \CV^\cM{}_\alpha\,\CV^\cM{}_\beta 
\;,
\ea
equals a constant.

\section{Gauged isometries}
\label{section:gauging}

In this section we elevate a subgroup of the isometries to a local
symmetry by making the parameters spacetime dependent. With increasing
$N$, supersymmetry then poses severe constraints on the possible
choices of gauge groups.

\subsection{Gauge group and embedding tensor}

A subgroup of isometries (possibly extended with R-symmetry
transformations for $N=2,4$) can be encoded in an embedding tensor
$\Theta_{\cM\cN}$ which defines the Killing vectors that generate the
gauge group by
\be
\label{subgroup}
X^i = g\,\Theta_{\cM \cN}\,\Lambda^\cM(x)\, X^{\cN \,i}\,,
\ee
with spacetime dependent parameters $\Lambda^\cN(x)$ and a gauge coupling
constant $g$. Unless the gauge group coincides with the full group of
isometries, the embedding tensor acts as a projector which
reduces the number of independent parameters. 
In order that this subset of Killing vectors generates a group,
$\Theta_{\cM\cN}$ must satisfy the following condition,
\be
\label{embedding-cond}
\Theta_{\cM\cP} \,\Theta_{\cN\cQ}\,f^{\cP\cQ}{}_\cR = \hat
f_{\cM\cN}{}^\cP\, \Theta_{\cP\cR}\,,
\ee
for certain constants $\hat f_{\cM\cN}{}^\cP$, which are subsequently
identified as the structure constants of the gauge group. 
One can verify that the validity of the Jacobi
identity for the gauge group structure constants follows directly from
the Jacobi identity associated with the full group of isometries.  For
a semi-simple gauge group the embedding tensor is simply the sum of
the Cartan-Killing forms of the different group factors with different
and a priori unrelated coupling constants. The embedding tensors of
non-semisimple gauge groups may take more complicated
forms~\cite{FiNiSa03}. 

The next step is to introduce gauge fields $A_\mu^\cM$ corresponding
to the gauge group parameters $\Lambda^\cM(x)$ and include them
into the definition of the covariant derivatives. For example, we have
\be
\CD_\mu\phi^i = \partial_\mu \phi^i +g\, \Theta_{\cM\cN} \,A^\cM_\mu \,
X^{\cN\,i} \,,
\label{covscalar}
\ee
which transforms under local isometries according to 
\be
\CD_\mu\phi^i \to \CD_\mu\phi^i + g\, \Theta_{\cM\cN} \,\Lambda^\cM
\,\partial_j X^{\cN\,i}\; D_\mu\phi^j\,,
\ee
provided we assume the gauge fields transformations, 
\be
\Theta_{\cM\cN} \,\delta A_\mu^\cM =  \Theta_{\cM\cN} \left(-\partial_\mu
\Lambda^\cM +g\, \hat f_{\cP\cQ} {}^{\cM}
\,A_\mu^\cP\,\Lambda^\cQ  \right)\,.
\ee
The covariant field strengths follow from the commutator
of two covariant derivatives, {\it e.g.}, 
\be
  [{\cal D}_\mu, {\cal D}_\nu]\, \phi^i =
g \,\GTh_{\cM\cN}\, F_{\mu\nu}^\cM \, X^{\cN i} \;,
\label{DDF}
\ee
and take the form 
\ba
\label{field-strength}
\Theta_{\cM\cN}\,F_{\mu\nu}^\cM = \Theta_{\cM\cN}\Big( 
\dd_{\mu}A_{\nu}^\cM -\dd_{\nu}A_{\mu}^\cM - 
g\, \hat f_{\cP\cQ}{}^{\cM} \, A_{\mu}^\cP A_{\nu}^\cQ\Big) 
\;.
\ea
The gauge transformations on the fermion fields follow from
\eqn{iso-trans2} upon substitution of~\eqn{subgroup}. From this we
derive the covariant derivatives for the spinor fields,   
\begin{eqnarray}
\CD_\mu \psi^I_\nu &=& \left( \partial_\mu +\ft12
\omega^a_\mu \gamma_a \right ) \psi^I_\nu  +
\dd_\mu\phi^i Q_i^{IJ} \,\psi^J_\nu
+ g\, \GTh_{\cM\cN} A_\mu^\cM\,\CV^{\cN\,IJ} \,\psi^J_\nu \;,
\non[1ex]
{\cal D}_\mu \chi^{iI} &=& 
\left( \partial_\mu +\ft12 \omega^a_\mu \gamma_a \right)\chi^{iI} 
+\partial_\mu \phi^j \left( \Gamma^i_{jk}\, \chi^{kI} +
Q_j^{IJ} \chi^{iJ} \right)\non[.5ex]
&&{}+ g\, \Theta_{\cM\cN} A_\mu^{\cM} 
\left( \delta^i_j\, \CV^{\cN\,IJ} -\delta^{IJ}g^{ik}\,
D_k\CV^\cN{}_j 
\right) \chi^{jJ} \;.
\label{covfermions}
\end{eqnarray}
In view of the commutator \eqn{QG-comm}, the covariant derivative on the
supersymmetry parameter acquires also an additional covariantization, 
\be
\CD_\mu \epsilon^I = \left( \partial_\mu +\ft12
\omega^a_\mu \,\gamma_a \right ) \epsilon^I   + 
\dd_\mu\phi^i \,Q_i^{IJ} \,\epsilon^J + g \, 
\GTh_{\cM\cN} \,A_\mu^\cM\,\CV^{\cN\,IJ} \,\epsilon^J \;.
\label{covepsilon}
\ee
In this section we only make use of the previous results \eqn{DXS}
that apply to arbitrary $N>0$.

The extra minimal couplings \Ref{covscalar} induce modifications of
the supersymmetry variations and of the Lagrangian. As long as we are
dealing with first derivatives, these new couplings do not lead to
complications as they are controlled by gauge covariance. However,
commutators of the new covariant derivatives lead to new (covariant)
terms proportional to the field strength \eqn{field-strength}. These
terms, which cause a violation of supersymmetry, take the form,
\ba
\Gd\CL &=& -\ft{1}{2}\I\,g\,\GTh_{\cM\cN} F_{\nu\rho}^{\cN}\,
\Ge^{\mu\nu\rho}\,\left( \CV^{\cN\, IJ}\,\Bpsi{}_\mu^I\Ge^J +
\ft{1}{2}\,\CV^{\cN}{}_i\, \Bchi^{i I}\Gg_\mu\Ge^I \right)
 \;.
\nn
\ea
They are cancelled by introducing a Chern-Simons term for the vector
fields, 
\ba
\CL_{\rm CS} &=& \ft{1}{4}\,\I g\,\varepsilon^{\mu\nu\rho} \,
A_\mu^{\cM} \,\GTh_{\cM\cN} \Big( \dd_{\nu} A_{\rho}^\cN - 
\ft13 g\, \hat f_{\cP\cQ}{}^{\cN} \, A_{\nu}^\cP A_{\rho}^\cQ\Big)  \;,
\label{LCS}
\ea
provided the embedding tensor $\Theta_{\cM\cN}$ is symmetric and
provided we assume the following supersymmetry transformations,
\ba
\Theta_{\cM\cN}\,
\Gd A^\cM_\mu &=& \Theta_{\cM\cN}\, \Big[
2\,\CV^{\cM\,IJ}\, \Bpsi{}_\mu^I\Ge^J
+\CV^\cM{}_{i} \, \Bchi^{i I}\Gg_\mu\Ge^I \Big]  \;.
\label{susyA}
\ea
At this point one can also derive the field
equation for the vector fields, which reads,
\ba
\label{fe-vector}
\Theta_{\cM\cN}\Big[ \hat F^\cM_{\mu\nu}+ 2 \,\I e
\varepsilon_{\mu\nu\rho} \,\hat 
D^\rho \phi^i\,\CV^\cM{}_i  + \ft1{12}\bar\chi^{iI} \gamma_{\mu\nu}
\chi^{jJ} \,
(g_{ij}\,\CV^{\cM IJ} -\delta^{IJ}\,D_i\CV^{\cM}{}_j ) \Big] =0\,,
\ea
where $\hat F^\cM_{\mu\nu}$ denotes the supercovariant curvature. 

The embedding tensor is a gauge group invariant element of ${\sf
Sym}(\mathfrak{g}\otimes\mathfrak{g})$ and therefore satisfies $\hat
f_{\cM\cN}{}^{\!Q}\, \Theta_{\cQ\cP}+ \hat f_{\cM\cP}{}^{\!Q}\,
\Theta_{\cQ\cN}=0$, which implies
\ba
  \Theta_{\cP\cL}\left(f^{\cK\cL}{}_\cM \Theta_{\cN\cK}
+ f^{\cK\cL}{}_\cN \Theta_{\cM\cK}\right) &=& 0 \;. 
\label{subgrouptheta}
\ea
Consequently, the structure constants of the gauge group can be
expressed as $\hat{f}_{\cM\cN}{}^\cP =
\GTh_{\cM\cQ}\,f^{\cP\cQ}{}_\cN$, as they satisfy \eqn{embedding-cond}
by virtue of \eqn{subgrouptheta}.

We define the so-called $T$-tensor
(originally introduced in higher-dimensional
supergravity~\cite{dWiNic82}) as the image of $\GTh$ under
the map $\CV$ from \Ref{Vmap}, {\it i.e.}
\ba
\begin{array}{rclrcl}
T^{IJ,KL} &\equiv& \CV^{\cM\,IJ}\GTh_{\cM\cN}\CV^{\cN\,KL}
\;,\qquad
&
T^{IJi} &\equiv&  \CV^{\cM\,IJ}\GTh_{\cM\cN}\CV^{\cN\,i}
\;,
\\[1ex]
T^{ij} &\equiv& \CV^{\cM\,i}\GTh_{\cM\cN}\CV^{\cN\,j}
\;,
&
T_\Ga{}^i &\equiv&  \CV^{\cM}{}_{\Ga}\GTh_{\cM\cN}\CV^{\cN\,i}
\;,
\\[1ex]
T_{\Ga\Gb} &\equiv& \CV^{\cM}{}_{\Ga}\GTh_{\cM\cN}\CV^{\cN}{}_{\Gb}
\;,
&
T^{IJ}{}_\Ga &\equiv&  \CV^{\cM\,IJ}\GTh_{\cM\cN}\CV^{\cN}{}_{\Ga}
\;. 
\end{array}
\label{T-tensors}
\ea 
The $T$-tensor components that carry indices $\alpha,\beta$ do not
play an important role, as they do not appear directly in the
Lagrangian and transformation rules. For the case of $N=2$ these
components are not defined. From~\Ref{subgrouptheta}, \eqn{cov-CV} and
\eqn{fABC}, it readily follows that the $T$-tensor transforms
covariantly under the gauged isometries. Furthermore, we note the
following identities,
\ba
\label{TT-rel}
D_{(i} T_{jk)} &=& 0\,, \nonumber \\
D_{(i}T^{IJ}{}_{j)} &=& \ft12 T_{k(i}\, f^{IJk}{}_{j)} \,, \nonumber \\
D_i T^{IJ,KL} &=& \ft12  f^{IJ}_{ij} \, T^{KLj}  + \ft12 f^{KL}_{ij} \,
T^{IJj} \,.
\ea
The covariance under the gauged isometries also allows the derivation
of identities quadratic in the $T$-tensors. Two examples of such 
identities are,
\ba
T^{MN \,i}\,T^{KL\, j} \,f_{ij}^{IJ} +T^{MN \,i}\,T^{IJ\, j}
\,f_{ij}^{KL}  &=& 4\,T^{MN,P[I}\, T^{J]P,KL} + 4\,T^{MN,P[K}\,
T^{L]P,IJ} \;, \nonumber \\ 
T^{ki}\,T^{KL\, j} \,f_{ij}^{IJ} +T^{ki}\,T^{IJ\, j}
\,f_{ij}^{KL}  &=& 4\,T^{P[I\,k}\, T^{J]P,KL} + 4\,T^{P[K\,k}\,
T^{L]P,IJ} \;. 
\ea

\subsection{Constraints from supersymmetry}

The supersymmetry variations of the vector fields in \Ref{covscalar},
\Ref{covfermions} cause additional supersymmetry variations of order
$g$. The variations linear in the spinor fields are
\ba
\Gd \CL &=& - eg\,\GTh_{\cM\cN} 
\left(
2\,\CV^{\cM\,IJ}\,\Bpsi{}_\mu^I\Ge^J
+ \CV^{\cM}{}_{i}\,\Bchi^{i I}\Gg_\mu\Ge^I \right)
\CV^{\cN}{}_j \,\CD^\mu\phi^j \;.
\label{new2}
\ea
They should be cancelled by introducing mass-like terms,
\ba
\CL_{g} &=&e g\Big\{
\ft12  A_1^{IJ}\,\Bpsi{}^I_\mu\,\Gg^{\mu\nu}\,\psi^J_\nu\, +
 A^{IJ}_{2\,j}\,\Bpsi{}^I_\mu\,\Gg^\mu \chi^{j J} +
\ft12  A_{3\,}{}_{ij}^{IJ}\, \Bchi^{i I}\chi^{j J}\Big\} 
\;,
\label{LY}
\ea
accompanied by additional modifications of the supersymmetry
transformation rules 
\ba
\Gd_g \psi^I_\mu = g\,A_1^{IJ} \Gg_\mu\,\Ge^J \;,
\qquad
\Gd_g \chi^{iI} = -  g N \, A_{2}^{JiI} \,\Ge^J
\;.
\label{fshifts}
\ea
Obviously the tensors $A_1$ and $A_3$ are symmetric,
$A_1^{IJ}=A_1^{JI}$,
$A_{3\,}{}_{ij}^{IJ}=A_{3\,}{}_{ji}^{JI}$. Furthermore, in view of
\Ref{projchi}, $A_2$ and $A_3$ are subject to the constraints, 
\ba
\label{proj-constraints}
\bbP_I^J{}_i^j\, A^{K\,J}_{2\,j} &=&  A^{KI}_{2\,i} \;,\qquad
\bbP_I^J{}_i^j\,A_{3\,}{}^{JK}_{jk} ~=~ A_{3\,}{}^{IK}_{ik}
 \;.
\ea
The variations of \Ref{LY} and the additional variations \Ref{fshifts}
of the original Lagrangian together cancel the terms \Ref{new2},
provided that $A_{2\,i}^{I\,J}$ and $A_3{}^{IJ}_{ij}$ take the
following form,
\ba
A_{2\,i}^{I\,J} &=& 
{1\over N} \Big\{ D_i A_1^{IJ} + 2\,T^{IJ}{}_i 
\Big\}  \;,
\non[1ex]
A_{3\,}{}^{IJ}_{ij}&=&
{1\over N^2} \Big\{ -2\,D_{(i}D_{j)}A_1^{IJ} + g_{ij}\,A_1^{IJ} +
A_1^{K[I} \,f_{ij}^{J]K} \non 
&& \hspace{9mm} 
+2\, T_{ij} \, 
\delta^{IJ}  
- 4\, D_{[i} T^{IJ}{}_{j]}-   2\, T_{k[i} \,f^{IJk}{}_{j]}  
\Big\}  \;.
\label{id1}
\ea
Here $A_3$ has the required symmetry structure. For the
convenience of the reader we also give the (dependent) result,
\ba
\label{D-A2}
D_i A_{2j}^{I\,J} &=& \ft12\,g_{ik}A_1^{IK}\,\bbP^K_J{}^k_j
-\ft{1}2N\, A_{3\,}{}^{IJ}_{ij}
+ T_{ik}
\,\bbP^I_J{}^k_j \;.
\ea
In addition, we need to ensure that both $A_2$ and $A_3$ as defined in
\Ref{id1} satisfy the projection constraints
\eqn{proj-constraints}. In view of \Ref{D-A2}, it is sufficient
to impose this constraint on $A_2$, which implies the following two
equations,
\ba
f^{K(I \,j}{}_i\,D_j A_1^{J)K} + (N-1) \,D_i A_1^{IJ} + 2\,D_i
T^{IK,JK} &=& 0\;, \nonumber\\[1ex]
f^{K[I \,j}{}_i\,(D_j A_1^{J]K}+2\,T^{J]K}{}_j)  - 2(N-1) \,T^{IJ}{}_i 
&=& 0\;. 
\label{proj-2}
\ea
These have several consequences.
By iterating the first equation ({\it i.e.}\ by resubstituting the
result for $D_iA_1$), we derive
\ba
\label{T+A}
f^{KI \,j}{}_i\,D_j (4\,T^{JL,KL} + (N-2) A_1^{JK})&& \nonumber \\
 - D_i(4\,T^{IL,JL} + (N-2) A_1^{IJ}) &=& 
  f^{IJ \,j}{}_i\,D_j A_1^{KK}+ \delta^{IJ}\, D_i A_1^{KK} \;.
\ea
This result can be combined with the second equation \eqn{proj-2} to
eliminate $A_1^{IJ}$ and to find a linear constraint for the
components $T^{IJ}{}_i$ of the $T$-tensor
\ba
\label{T3-id}
(N-4)\, T^{IJ}{}_i + 2\,f^{K[I}_{ij} \,T_{\vphantom{j}}^{J]K\,j} 
- {1\over N-1}
\,(f^{IJ}\,f^{KL})_{ij} \,T^{KL\,j} &=& 0\;.
\ea
Applying this constraint to the combination
$f^{IJ}\,T^{KL}+f^{KL}\,T^{IJ}$, the resulting equation may be
integrated to
\ba
(N-2) \left(T^{IJ,KL}- T^{[IJ,KL]}\right) 
- 4\,\delta\oversym{^{I[K}\,T^{L]M,JM}\!\!\!}\;\, 
+\frac{2}{N\mis1}\,\delta^{I[K}\delta^{L]J}\,T^{MN,MN} = 0 \;.
\quad
\label{con0}
\ea
A priori, this equation holds up to a covariantly constant term.
Because of (\ref{fQ}), covariantly constant terms cannot exist,
unless they are SO$(N)$ invariant and therefore constant.  However,
the above equation does not contain a singlet contribution so that it
is in fact exact for any $N$.

Vice versa, one can show that the covariant derivative of \Ref{con0}
implies \Ref{T3-id}, such that these two equations are in fact
equivalent. It is not straightforward to disentangle various
independent equations, due to the nontrivial properties of the $\ft12
N(N\mis1)$ almost-complex structures $f^{IJ}$. For example, the
following equation is not independent,
\ba
(N\mis8)\,D_iT^{[IJ,KL]} - {1\over N\mis1} (f^{[IJ}f^{KL]})_{ij}\, D^j
T^{MN,MN} + 5 (f^{M[I}f^{JK})_{ij} T^{L]Mj} = 0\;.\quad
\ea
One can systematize
this analysis by employing a set of ${\rm SO}(N)$ projection operators, 
as we briefly sketch in appendix~\ref{SONprojectors}. 

Now we use \eqn{T3-id} to rewrite the $f^{KI}\,D T^{JL,KL}$ term in
\eqn{T+A}. Combining the result with \eqn{proj-2} to remove the
$f^{KI}\,D A_1^{JK}$ terms, we may integrate the resulting equation to
obtain
\ba
4\,T^{IL,JL} + (N-2) A_1^{IJ}- {2\over N\mis1}\, T^{MN,MN}\,\delta^{IJ} 
&=& \mu \, (N-2)\, \delta^{IJ} \;,
\label{A1T}
\ea
with an as yet undetermined constant $\mu$. Putting things together,
we have shown that 
supersymmetry at linear order in $g$ determines the tensors $A_1$,
$A_2$, $A_3$ according to \Ref{id1}, \Ref{A1T} in terms of the
$T$-tensor \Ref{T-tensors} while the latter satisfies the (equivalent)
constraints \Ref{T3-id}, \Ref{con0}.

Before proceeding to the remaining terms in the action and
transformation rules, we take a brief look at the supersymmetry
algebra. The supersymmetry commutator leads to a covariantized
translation, and a supersymmetry and Lorentz transformation with
parameters proportional to $\chi^2$. When switching on the gauge
coupling, there is an extra Lorentz transformation, but more
importantly, also a local isometry with parameter given by
\be 
{[}\delta(\epsilon_1), \delta(\epsilon_2)] = \delta_{\rm G}(2\,
\CV^{\cM\,IJ} \,\bar\epsilon_2^I\epsilon_1^J) + \cdots\,. 
\ee
The supersymmetry established so far guarantees the closure of the
algebra to that order, except for the gauge fields which appear
multiplied by a coupling constant. Their closure (up to the field
equations \eqn{fe-vector}) implies another constraint, namely,
\ba
\GTh_{\cM\cN}(2\, \CV^{\cN\, K(I} \, A_1^{J)K}+ \CV^{\cN i}
D_i A_{1}^{IJ}) &=&0  \;. 
\label{id2}
\ea
It implies that the function $A_1$ is gauge covariant; in
particular, its trace is invariant, {\it i.e.}\ $\Theta_{\cM\cN}
\CV^{\cM i}\,D_iA^{II}=0$. This is in agreement with equation
\Ref{A1T} since we have already proven that the
$T$-tensors are gauge covariant. Moreover, equations \Ref{id1} show
that the tensors $A_2$ and $A_3$ are covariant as well, because they
depend on the $T$-tensors and $A_1$ and covariant derivatives
thereof. Again we can derive certain identities from \eqn{id2} that
involve some of the $T$-tensors and $A_1$, such as
\ba
T^{IJ\,i}\,D_iA_1^{KL} 
+ 2\, T_{\vphantom{1}}^{IJ,M(K} A_1^{L)M}&=& 0\;,\nonumber\\ 
T^{ij}\,D_jA_1^{KL} 
+ 2\, T_{\vphantom{1}}^{M(K\,i} A_1^{L)M}&=& 0\;.
\label{extraid}
\ea

In order to preserve supersymmetry to order $g^2$ one determines the
corresponding variations linear in $\psi^I_\mu$ and $\chi^{iI}$. They
reveal the need for a (gauge invariant) scalar potential in the
Lagrangian, 
\ba
\CL_{g^2} ~=~ -eV &\equiv& 
{4\, e g^2\over N}  \left( A_1^{IJ}A_1^{IJ} -\ft12 N\,
g^{ij}\, A_{2i}^{I\,J} A_{2j}^{I\,J} \right) \non[.5ex]
&=& 
{4\, e g^2\over N^2}  \left(N\,  A_1^{IJ}A_1^{IJ} -\ft12 
g^{ij}\, D_iA_1^{IJ} \,D_j A_1^{IJ} -2\,g^{ij} \, T^{IJ}_i\,T_j^{IJ}
\right)\;. 
\label{potential}
\ea
We note that the variation of the scalar potential is given by
\ba
\dd_i \CL_{g^2} &=& e\,g^2 \Big(
3\,A_1^{IJ}\,A_{2i}^{I\,J} + N\, A_3{}^{IJ}_{ij}\,A_{2}^{IjJ} \Big) 
 \;,
\label{varpotential}
\ea
by virtue of \eqn{id1}, \eqn{D-A2} and \eqn{id2}. In order that all
supersymmetry variations of the potential cancel, the following two
quadratic equations must be satisfied,
\ba
2\,A_1^{IK}A_1^{KJ} - N \, A_{2}^{IiK}\,A_{2i}^{J\,K}
&=& {1\over N}\,\Gd^{IJ} \left(
2 \,A_1^{KL}\,A_1^{KL} -  N\,A_{2}^{KiL}A_{2i}^{KL}
\right) \;,
\non[1ex]
3\,A_1^{IK}\,A_{2j}^{K\,J} 
+ N\,g^{kl}\,A_{2k}^{IK}\,A_3{}^{KJ}_{lj} &=&
\bbP_J^I{}_j^i \left(
3\,A_1^{KL}\,A_{2i}^{KL} + N\,g^{kl}\,A_{2k}^{LK}\,A_3{}^{KL}_{li} 
\right)\;.
\label{id3}
\ea
It can be shown after some computation that these relations are a
direct consequence of \Ref{extraid} upon using \Ref{id1} and
\Ref{A1T} and require the integration constant $\mu$ in the latter
equation to vanish.

What remains is to analyze the supersymmetry variations cubic in the
fermion fields. Their cancellation depends almost entirely on the
results presented so far. {\it E.g.} supersymmetry
variations proportional to $\psi^3$ cancel provided that
\ba
  \delta\oversym{_{\vphantom{1}}^{I[K} A_1^{L]J}\;}
  &=& - T^{IJ,KL} + T^{[IJ,KL]} \;,
\label{cub1}
\ea
which is in agreement with equation (\ref{A1T}). The variations that
are proportional to $\chi\psi^2$ cancel by virtue of \Ref{id1}. We
have not verified the cancellation of the supersymmetry variations
proportional to $\chi^2\psi$ and $\chi^3$, but we expect that
all these terms vanish by means of the constraints derived so far. In
the case of the maximal $N=16$ theory this has been verified
explicitly~\cite{NicSam01a}.

The central result of this
paper, is that a gauge group $G_0$ with a gauge invariant embedding tensor
$\Theta_{\cM\cN}$ describing the minimal couplings according to
\Ref{covscalar}, is consistent with supersymmetry if and only if the
associated $T$-tensor \Ref{T-tensors} satisfies the constraint, 
\ba
T^{IJ,KL} &=& T^{[IJ,KL]} 
- \frac4{N\mis2}\,\delta\oversym{^{I[K}\,T^{L]M,MJ}\;} 
- \frac{2\,\delta^{I[K}\delta^{L]J}}{(N\mis1)(N\mis2)}\,T^{MN,MN} \;.
\label{relT}
\ea
{}From this constraint all further consistency conditions can be
derived. The Lagrangian is modified by a Chern-Simons term \Ref{LCS},
mass-like fermionic terms \Ref{LY} and a scalar potential 
\Ref{potential} and the fermions have additional supersymmetry
variations \eqn{fshifts}. Note that the constraint (\ref{relT}) is
well-defined even for $N=1$ and $N=2$, but degenerates into an identity. 
The consistency constraint \Ref{relT} has a simple group theoretical
meaning in $\SO{N}$: denoting the irreducible parts of $T^{IJ,KL}$
under $\SO{N}$ by 
\ba
\Yboxdim6pt
{\young(\hfil,\hfil)} \;\times _{\rm sym} \, 
{\young(\hfil,\hfil)}
~\;=\;~ 1 \;\;+\;\;
{\young(\hfil\hfil)} \;\;+\;\;
{\young(\hfil\hfil,\hfil\hfil) }
\;\;+\;\; 
{\yng(1,1,1,1) }\;\; ,
\label{IJKL}
\ea
with each box representing a vector representation of
$\SO{N}$,\footnote{
  We use the standard Young tableaux for the orthogonal groups; {\it i.e.}\
  the four representations in the decomposition 
  \Ref{IJKL} are of dimension 1, $\ft12 N(N\pls1) - 1$,
  $\ft1{12}N(N\mis3)(N\pls1)(N\pls2)$, and ${N}\choose{4}$,
  respectively. For $N=8$, the last representation is 
  reducible, but this does not affect the argument here.} 
equation \Ref{relT} expresses that
\ba
\Yboxdim3pt
\bbP\!_{\atop{}{\yng(2,2) }}\; 
T^{IJ,KL} &=& 0\;.
\label{con88}
\ea

\mathon
\section{The theories with $N\le4$}
\mathoff
\mathon
\subsection{$N=1$}
\mathoff

In this case, the target space is a Riemannian
manifold of arbitrary dimension $d$. The consistency conditions for
the gauged theory simplify considerably; in particular the quadratic
constraints (\ref{id3}) become identities.

The tensor $A_1$ has just one component, which defines a function $F$
on the target space. According to
(\ref{id2}) $A_1$ is gauge invariant, and so is $F$,
\be
  \GTh_{\cM\cN}X^{\cN i}\dd_i F = 0 \;.
  \label{N1C}
\ee
Reading off the values for $A_2$ and $A_3$ from (\ref{id1}), we obtain 
\ba
    A_1 &=& F\;, \qquad
    A_{2\,i} ~=~ \dd_i F \;, \qquad
    A_{3\,ij} ~=~ g_{ij} F -2\, D_i\dd_j F +
    2\, T_{ij} \;,
  \label{N1A123}
\ea
with $T_{ij}=X^{\cM}_i\, \GTh_{\cM\cN} X^{\cN}_j\,$.

As a consequence of (\ref{N1C}), any subgroup of isometries can be
gauged (for example, by choosing a constant function $F$). The
gravitino $\psi_\mu$ is never charged under the 
gauge group, as can be seen directly from (\ref{covfermions}), and the
gauging is restricted to the matter sector. The scalar potential $V$
(\ref{potential}) is given by
\be
  V = - g^2 \left( 4 \,F^2 - 2\, g^{ij}\,\dd_iF\dd_jF \right) \;,
\label{N1potential}
\ee
{\it i.e.}\ the function $F$ serves as the {\em real
superpotential}. Supersymmetry (in a maximally symmetric spacetime)
requires the vanishing of $A_2$ ({\it c.f.}~\Ref{susy-stat1} below),
so that the stationary points of $F$ define (anti-de Sitter)
supersymmetric ground states.

There exist deformations of the original theory that are not induced
by gaugings. They correspond to $\GTh_{\cM\cN}=0$ and $F\neq0$ and are
described by the Lagrangian (\ref{L}), together with the mass-like
terms (\ref{LY}), subject to (\ref{N1A123}), and with the scalar
potential (\ref{N1potential}).

\mathon
\subsection{$N=2$}
\mathoff
The target space is now a K\"ahler manifold.  Some (partial) results
for abelian gaugings have already been obtained in
\cite{DKSS99,AboSam01}. As it turns out, any subgroup of the
invariance group can be gauged. These gaugings share some features
with the $N=1$ gaugings of four-dimensional
supergravity~\cite{BagWit82,Bagg83,CFGP83}.

Many of the quantities introduced above simplify considerably. It is
therefore convenient to introduce the notation
\ba
  f   &=& f^{12} \;, \qquad
  Q_i ~=~ Q^{12}_i \;, \qquad
  \mathcal{V} ~=~ \mathcal{V}^{12}\;,  \non 
  T_i&=& T^{12}{}_{\!i} \;, \qquad  
  T= T^{IJ,IJ}= 2\, T^{12,12} \;.
\ea
To avoid confusion, we keep using the notation $\Lambda^{12}$
and ${\mathcal{S}}^{12}$ for the parameters of the $\SO2$
transformations. Further, we have
\ba
    \partial_i Q_j - \partial_j Q_i &=& \ft12 f_{ij} \;,
\qquad
    D_i(\Gamma) f_j{}^k ~=~ 0 \;,
\label{partial-Q} 
\ea
where $\Gamma^k_{ij}$ is the Christoffel connection.  For a K\"ahler
target space it is convenient to decompose the $d$ real
fields into $d/2$ complex ones and their complex conjugates,
$\phi^i \to (\phi^i, \Bphi{}^\Bi)$ in a basis where
$f_i{}^j=\I\,\delta_i^j$, $f_\Bi{}^\Bj=-\I\,\delta_\Bi^\Bj$.  From the
fact that $f$ is hermitian, it follows that only the components
$g_{i\bar\jmath} = g_{\bar\imath j}$ are non-zero, and therefore
$f_{i\bar\jmath} = \I g_{i\bar\jmath} = - f_{\bar\jmath i}$.  The fact
that $f$ is covariantly constant then leads to
\be
  \partial_i g_{j\bar k} = \partial_j g_{i \bar k} \,,
\ee
which implies that the metric can locally be written as
$g_{i\bar\jmath} = \partial_i \partial_{\bar\jmath} K$, 
where $K(\phi,\bar \phi)$ is the K\"ahler potential. Furthermore
\eqn{TT-rel} implies,
\ba
T_i = \ft12 \I \,\partial_i T\;, \qquad T_{ij} = -D_i\partial_jT\;. 
\ea
The projectors defined in (\ref{projchi}), decompose into a
holomorphic and an anti-holomorphic component,
\be
  \mathbb{P}^{Ii}_{Jj} = 
  \ft12\, \delta^i_j \, 
\bigl(\delta^{IJ} + \I \epsilon^{IJ} \bigr) \,,
  \quad
  \mathbb{P}^{I\bar\imath}_{J\bar\jmath} = 
  \ft12\, \delta^{\bar\imath}_{\bar\jmath} \,
\bigl(\delta^{IJ} 
  - \I \epsilon^{IJ} \bigr) \,.
\ee
The K\"ahler potential $K$ is defined up to K\"ahler transformations,
\be
  K(\phi,\bar\phi) \to K(\phi,\bar\phi) + \Lambda(\phi) +
  \bar\Lambda(\bar\phi) \,.
  \label{kahler-trans}
\ee
A solution of (\ref{partial-Q}) is provided by
\be
  Q_i = -\ft14{\I} \partial_i K \;,
  \qquad
  Q_{\bar\imath} = \ft14{\I}  \partial_{\bar\imath} K \,.
  \label{Q-K}
\ee
This solution is not unique as it is subject to field-dependent gauge
transformations. By adopting (\ref{Q-K}) we have removed this gauge
freedom, but the K\"ahler transformations now act on $Q$ in the form
of a field-dependent $\SO2$ gauge transformation with parameter
\begin{equation}
  \Lambda^{12}(\phi,\bar\phi) = \ft14{\I} \bigl( \Lambda(\phi)
  -\bar{\Lambda}(\bar\phi)\bigr) \,.
  \label{lambda-12}
\end{equation} 
Consequently, all quantities transforming nontrivially under $\SO2$
become now subject to K\"ahler transformations induced by
(\ref{lambda-12}). Note that transformations where $\Lambda$ equals
an imaginary constant, correspond to $\SO2$ transformations acting
exclusively on the fermions and not on the K\"ahler potential. These
transformations constitute an invariance group of the ungauged
Lagrangian and they are in the center of the full group of combined
isometries and $\SO2$ transformations of the fermions.

According to (\ref{isomf}) only holomorphic isometries of the target
space can be extended to symmetries of the Lagrangian. Such
isometries, parameterized by Killing vector fields $(X^i,X^{\bar
\imath})$, preserve both the metric and the complex structure, {\it
i.e.}, 
\ben
  \mathcal{L}_X g = \mathcal{L}_X f =0 \,.
\een
The invariance of the complex structure implies that $X^i$ and
$X^{\bar\imath}$ must be holomorphic and anti-holomorphic,
respectively.  The invariance of the metric gives rise to the Killing
equations
\ben
  D_i X_{\bar\jmath} + D_{\bar\jmath} X_i = 0 \;,\qquad
  D_i X_{j} + D_{j} X_i = 0 \,.
\een
Because of the holomorphicity of $X^i$, the second condition is
automatically satisfied, whereas the first condition implies that the
K\"ahler potential remains invariant under the isometry up to a
K\"ahler transformation. We write this special K\"ahler transformation
in terms of a holomorphic function~${\mathcal{S}}(\phi)$, {\it
i.e.}, 
\begin{equation}
  \delta K(\phi,\bar\phi) = - X^i\,\partial_i K -
  X^{\bar\imath}\,\partial_{\bar\imath}  K  = 
4 \I \,\bigl( {\mathcal{S}} -\bar{\mathcal{S}} \bigr)  \,.
  \label{special-kahler-trans}
\end{equation}
According to (\ref{DXS}), the function $\mathcal{V}$, defined
by 
\ben
  \mathcal{V} = X^i Q_i + X^{\bar\imath} Q_{\bar\imath} +
{\mathcal{S}}^{12} 
  = -\ft14{\I} X^i \partial_iK + \ft14{\I} X^{\bar\imath}
  \partial_{\bar\imath}K  + {\mathcal{S}}^{12}  \;,
\een
must satisfy the equation, 
\ben
  \partial_i \mathcal{V} = 
\ft12{\I}\, g_{i\bar\jmath} X^{\bar\jmath} \,.
\een
As the right-hand side can be written as a derivative, this equation
can now be solved and we obtain 
\be
  {\mathcal{S}}^{12}(\phi,\bar\phi) = {\mathcal{S}}(\phi) +
  \bar{\mathcal{S}}(\bar\phi) \,.
  \label{S-12}
\ee
Consequently we have,
\ba
  \mathcal{V} &=&  
-\ft14{\I} \bigl(X^i \partial_iK - X^{\bar\imath} 
  \partial_{\bar\imath}K \bigr)  + {\mathcal{S}} + \bar {\mathcal{S}}
~=~
  - \ft12{\I} X^i \partial_i K + 2\, \mathcal{S} \,.
\nn
\ea
For every generator $X^\cM$ of the isometry group we may thus identify
a holomorphic function~${\mathcal{S}}^\cM$, determined by
\Ref{special-kahler-trans} up to a
real constant. The particular transformation (which we  
denote with the extra label $\CM=0$)
\be
X^{0\, i}=0 \;,\qquad \CS^0 = \ft12  \;, \qquad \CV^0=1\;, 
\label{N2SO2}
\ee
constitutes a central extension of the isometry group and generates
the $\SO2$ R-symmetry group that acts exclusively on the fermions. The
closure relation~\Ref{SABC} yields,  
\ba
X^{\cM i}\partial_i{\mathcal{S}}^\cN -
X^{\cN i}\partial_i{\mathcal{S}}^\cM  &=& 
\sum_{\CK>0} f^{\cM\cN}{}_{\!\cK}\,\CS^\cK + f^{\cM\cN}{}_{\!0} 
\;.
\label{fMN0}
\ea
For a semi-simple isometry group, the $f^{\cM\cN}{}_0$ can be
absorbed by suitable constant shifts into the functions
$\mathcal{S}^\cM$ (or, equivalently, into  $\CV^\cM$).

It is always possible to gauge the R-symmetry group. In that
case we have a gauge field $A_\mu^0$ associated with ${\rm
SO}(2)$ transformations of the fermions and the $T$-tensor must be
manifestly invariant under this group. When the structure constants 
$f^{\cM\cN}{}_{\!0}$ do not vanish when projected onto the gauged
subgroup of the isometries, and cannot be absorbed by suitable shifts,
the R-symmetry group must be gauged. In 
that case it is more practical to choose a given set of functions
$\CV^\cM$ whose algebra will exhibit a certain central charge, without
trying to modify the structure constants by constants shifts. Instead
one can then vary the embedding tensor $\Theta_{\cM\cN}$. The reason is
that the Lagrangian is invariant under a change of basis in the Lie
algebra, according to $\mathcal{V}^\cM\to\mathcal{V}^\cM+c^\cM$,  
$A_\mu^\cM\to A_\mu^\cM+c^\cM A^0_\mu$, and 
\ba
\GTh_{\cM0}&\to&\GTh_{\cM0}-c^\cN \GTh_{\cN\cM} \;,
\qquad({\cal M}\not=0)\;,\non
\GTh_{00}&\to&\GTh_{00}- 2\,c^\cM \GTh_{\cM0}+ c^\cM
\GTh_{\cM\cN}c^\cN \;, 
\ea
where the $c^\cM$ are arbitrary constants with $c^0=0$. To see this
one observes that the combinations $\CV^{\cM}\Theta_{\cM\cN}\CV^{\cN}$,
$\CV^{\cM\,i}\Theta_{\cM\cN}\CV^{\cN}$  and
$\CV^{\cM\,i}\Theta_{\cM\cN}A_\mu^{\cN}$ remain invariant under the
combined substitutions. 

Let us further note that, given a
gauge group $G_0$ whose embedding tensor $\GTh$ satisfies
\Ref{subgrouptheta}, another solution to \Ref{subgrouptheta} can always be
obtained by the deformation
\ba
\GTh_{\cM 0} &\to& \GTh_{\cM 0} + \delta_\cM \;,
\label{shifttheta}
\ea
provided the generator $\delta_\cM X^\cM$ commutes with the gauge group
$G_0$. This is the three-dimensional analogue of the local version of
the Fayet-Iliopoulos mechanism in four-dimensional $N=1$
supergravity~\cite{Bagg83}. 

Let us now determine the various quantities involved. It is convenient
to decompose the tensor $A_1^{IJ}$ in terms of a singlet part
$A_1^{11}+A_1^{22}$ and a complex quantity
\ben
  e^{K/2}\,W ~\equiv~ \ft12 \bigl(A_1^{22}-A_1^{11}\bigr) 
+ \I A_1^{12} \;.
\een
K\"ahler transformations are induced by the $\SO2$ transformations
(\ref{lambda-12}),
\ben
  \delta \bigl( e^{K/2}\,W \bigr) = 
2 \I \,\Lambda^{12} \bigl( e^{K/2}\,W \bigr) \;.
\een
This implies that $W$ transforms under K\"ahler transformations
according to 
\begin{equation}
  \delta W  = - \Lambda(\phi)\, W \;.
  \label{delta-W}
\end{equation}
Imposing the equations (\ref{proj-2}) then leads 
directly to the following result,
\ben
  \partial_i \bigl(A_1^{11}+A_1^{22} + 2\, T\bigr) = 0 \,,
\qquad
  \partial_i\overline W =\partial_{\bar\imath} W = 0
\;.
\een
The function $W$ can be identified as the
{\em holomorphic superpotential}. Gauge covariance of $A_1$
imposes the additional relation
\ba
    \Theta_{{\cM}{\cN}} \bigl( X^{{\cN}i} D_i W 
    + 4 \I\, \mathcal{V}^\cN W \bigr)~=~\Theta_{{\cM}{\cN}} 
   \bigl(X^{{\cN}i} \partial_i W  
    + 4 \I\, \mathcal{S}^\cN W \bigr)
    ~=~ 0 \,,
\label{DW}
\ea
with the K{\"a}hler covariant derivative $D_i W \equiv \dd_i W + \dd_i
K\,W$, implied by (\ref{delta-W}). For nonvanishing $W$ the momentum
maps associated with the gauge group generators can be expressed in
terms of the superpotential and are proportional to the corresponding
Killing vectors. As a consequence, one may
verify from~\Ref{fMN0} that the structure constants
$f^{\cM\cN}{}_{\!0}$  vanish when projected onto the gauge group, {\it
i.e.}, $\GTh_{\cM\cK}\GTh_{\cN\cL}\,f^{\cK\cL}{}_{\!0} =0$. 
The gauging of the ${\rm SO}(2)$ R-symmetry group requires $W$ to
vanish (because $W$ transforms nontrivially under~${\rm SO}(2)$). 
Therefore nonzero $W$ implies that we have $\GTh_{\cM 0}=0=\GTh_{00}$.

The integration constant in $A_1^{11}+A_1^{22}$ is finally determined
by the quadratic constraints (\ref{id3}) provided that $W$ is
nonvanishing. The tensor $A_1^{IJ}$ is then given by
\be
A_1^{11}=-T-e^{K/2}\,\Re W \;,\quad\;
A_1^{22}=-T+e^{K/2}\,\Re W \;,\quad\;
A_1^{12}=A_1^{21}=e^{K/2}\,\Im W \;.
\ee
The tensor $A_2$ can be derived from (\ref{id1}) and its components
are given by
\ba
    A_{2\,i}^{1\,1} &=& -\I A_{2\,i}^{1\,2} ~=~
-\ft12 (\partial_i T   
    + e^{K/2} D_i W) \,, \non
    A_{2\,i}^{22} &=& \I A_{2\,i}^{21} ~=~
- \ft12 ( \partial_i T - e^{K/2} D_i W ) \,. 
\label{A-2-N-2}
\ea
Finally, the tensor $A_3$  
can be evaluated from (\ref{id1}); this leads to, for example,
\ba
A_3{}^{11}_{ij} &=& \ft14\,e^{K/2}\,D_iD_j W \;,
\non
A_3{}^{11}_{i\Bj} &=& \ft12 
T_{i\Bj} -\ft14\, g_{i\Bj}\,T + \ft12 \partial_i\partial_{\bar\jmath}T 
\;.
\ea

One can verify the consistency of these results by inserting $A_1$,
$A_2$ and $A_3$ into the quadratic constraints (\ref{id3}). Indeed, these 
cancel by virtue of (\ref{DW}). When $W=0$ the vanishing of the
integration constant in $A_1^{11}+A_1^{22}$ cannot be derived from
the quadratic constraints, but follows instead from the
identity~\Ref{cub1}. We note in passing that pure $N=2$ supergravity
(without gauging) can have a cosmological constant corresponding to a
constant $W$ and vanishing $T$. This implies that the gravitino mass
matrix is traceless. An alternative way to generate a cosmological
term in pure supergravity makes use of gauging the R-symmetry
group. In that case, $T$ equals a nonzero constant and $W=0$; the
gravitino mass matrix is then proportional to the identity.  

The scalar potential (\ref{potential}) of the gauged theory is
given by
\be
  V = - g^2 \Big(
  4\,T^2 - 4\,g^{i\Bi}\,\dd_iT\,\dd_\Bi T + 4\,e^K \, |W|^2 - 
  g^{i\Bi}\,e^K D_i W D_\Bi \overline{W}\Big) \;.
 \label{potentialN2}
\ee
Note that in three dimensions, the scalar potential is quartic in the
moment map $\mathcal{V}$, since the $T$-tensor is quadratic in
$\mathcal{V}$. This is in contrast with {\it e.g.}\ four dimensions, where
the scalar potential is quadratic in $\mathcal{V}$.

Analogous to the $N=1$ case, there are two kinds of supersymmetric 
deformations of the original theory. On the one hand, there are the
gaugings, which are completely characterized by an embedding tensor
$\Theta_{\cM\cN}$. The above analysis shows that there is no
restriction on the $T$-tensor, and therefore any subgroup of the
invariance group of the theory is an admissible gauge group, as long
as its embedding tensor satisfies~\Ref{subgrouptheta}. On the other
hand there are the deformations described by the holomorphic
superpotential $W$, which are not induced by a gauging. In case both
deformations are simultaneously present, their compatibility
requires~\Ref{DW}. 

\mathon
\subsection{$N=3$}
\mathoff
In this case the target space is a quaternionic manifold. The
condition \Ref{relT}, from which we have derived all other consistency
constraints, is identically satisfied, so that each subgroup of
isometries can be consistently gauged. The gauging follows uniquely
from the embedding tensor and there are no other deformations. In
particular, the scalar tensors $A_1$, $A_2$, $A_3$ are defined
by~\Ref{id1}, \Ref{A1T}, and the scalar potential is given by
\Ref{potential}.

\mathon
\subsection{$N=4$}
\mathoff
For $N=4$ the target space is locally 
a product of two quaternionic manifolds of dimension~$d_\pm$,
associated with the positive and negative eigenvalues of the tensor
$J$, whose real coordinates we denote by $\phi^i$, $\phi^{\Bi}$,
respectively. Because the almost-complex structures commute with $J$,
they decompose into two sets of three almost-complex structures
$f^\pm$;  the only nonvanishing components are
$f_{ij}^{+P}$, $f_{\Bi\Bj}^{-\bar P}$, 
where $\pm$ denotes the split according to ($P,\bar P=1,2,3$) 
\ba 
f^{+P} &\equiv& \ft12\,(J+ 1)\,f^P ~=~
\ft12 f^{P} - \ft14 \Ge^{PQR}\,f^{QR} \;,\non 
f^{-\bar P} &\equiv& \ft12\,(J- 1)\,f^P ~=~
-\ft12 f^{P} - \ft14 \Ge^{PQR}\,f^{QR} \;.
\label{PIJ}
\ea
The two sets of almost-complex structures satisfy the multiplication
rule, 
\ba
f^{+ P}\,f^{+Q}~=~- {\bf 1}\,\delta^{PQ} -\varepsilon^{PQR} f^{+ R}\;,  
\qquad
f^{-\bar P}\,f^{-\bar Q}~=~- {\bf 1}\,\delta^{\bar P\bar Q} -\varepsilon^{\bar
P\bar Q\bar R} f^{-\bar R}  \;.
\ea
There is a corresponding decomposition of the ${\rm SO}(4)$
R-symmetry group, 
\ba
\SO4 =  \SO3^+ \times \SO3^- \;.
\label{SO33}
\ea
Obviously the isometry group splits into the product of the 
isometry groups of the two subspaces whose generators we label by
$X^{{\cM} i}$ and $X^{{\bar \cM} \Bi}$, respectively. These $N=4$
three-dimensional gaugings share some similarities with the $N=2$ 
gaugings of four-dimensional
supergravity~\cite{dWLavP85,DAFeFr91,ABCD97a,ABCD97b}, although the
precise relation remains to be understood. Upon reduction to three
spacetime dimensions the special K\"ahler and
the quaternionic manifolds that describe the
interactions of the four-dimensional vector multiplets and hypermultiplets,
respectively, give rise to the two quaternionic spaces that span the
target space manifold \cite{CeFeGi89,JWKV97}. 

Let us first consider the nondegenerate case $d_+ d_- \not= 0$.
According to the discussion of equation~\Ref{S}, the quantities
$\CV^{\cM \,IJ}(\phi,X)$ decompose into two triplets denoted by ${\cal
V}^{{\cM}\,P}$ and ${\cal V}^{{\bar\cM}\,\bar P}$, which live in each
of the corresponding subspaces. The triplets $\CV$ are the momentum
maps associated with the isometries of the quaternionic
spaces~\cite{Gali87}.

A priori, the embedding tensor $\GTh_{\cM\cN}$ decomposes into
diagonal components, $\GTh_{\cM\cN}$ and $\GTh_{\bar\cM\bar\cN}$, and
off-diagonal components $\GTh_{\cM\bar\cN}$. The latter are, however,
severely constrained by the invariance 
condition \Ref{subgrouptheta},
\ba
\GTh_{\cP\cM}\,f^{\cM\cN}{}_{\cK}\,\GTh_{\cN\bar \cL} 
&=&
\GTh_{\cP\bar\cN}\,f^{\bar\cM\bar\cN}{}_{\bar\cL}\,\GTh_{\bar\cM\cK} 
\;,\non
\GTh_{\bar\cP \cM}\,f^{\cM\cN}{}_{\cK}\,\GTh_{\cN\bar \cL} 
&=&
\GTh_{\bar\cP\bar\cN}\,f^{\bar\cM\bar\cN}{}_{\bar\cL}\,\GTh_{\bar\cM\cK} 
\;.
\label{thetapm}
\ea
Under~\Ref{SO33}, the $T^{IJ,KL}$ component of the $T$-tensor
\Ref{T-tensors} takes values in the representations, 
\ba
T^{IJ,KL} \;:\quad 
{\bf (1,1)} + {\bf (1,1)} + {\bf (5,1)} + {\bf (1,5)} + {\bf
(3,3)}
\;.
\label{TIJKL4}
\ea
The constraint \Ref{relT} which is necessary and sufficient for the
existence of a supersymmetric gauging, implies the absence of the
${\bf (1,5)} + {\bf (5,1)}$ representation in this decomposition;
in the basis \Ref{PIJ} it implies
\ba
T^{P Q} &=& \ft13\,\delta^{P Q}\,T^{R R}
\;,\qquad\mbox{where}\quad
T^{P Q} ~=~
\CV^{\cM\,P}\GTh_{\cM\cN}\CV^{\cN\,Q}
\;,
\label{conN4}
\ea 
and correspondingly for $T^{\bar P\bar Q}$. The off-diagonal
components, $T^{P\bar Q}$, which are proportional to
$\GTh_{\cM\bar\cN}$, remain unconstrained. Unlike the cases $N<4$, it
is no longer possible to gauge any subgroup of the isometry group; the
consistency of the gauged theory depends on the condition~\Ref{conN4}
for the momentum maps $\CV^{\cM\,P}$ and $\CV^{\bar\cM\,\bar P}$ and
the gauge group invariant diagonal components $\GTh_{\cM\cN}$ and
$\GTh_{\bar\cM\bar\cN}$ of the embedding tensor, together with the
compatibility relation~\Ref{thetapm} for the off-diagonal components
$\GTh_{\cM\bar\cN}$ of the embedding tensor. For symmetric
quaternionic spaces there are convenient techniques for finding
admissible gauge groups, as we will discuss in the next
section. However, for non-symmetric~\cite{Alek75,dWiVPr92,Cort96} or
even non-homogeneous quaternionic spaces it remains to directly
analyze equation~\Ref{conN4} in order to determine the possible
solutions for~$\GTh_{\cM\cN}$.

Let us finally discuss the degenerate case in which one of the two
quaternionic manifolds vanishes, {\it i.e.}\ let us assume that
$d_-=0$. The Lagrangian \Ref{L} then admits an
additional global 
symmetry $SO(3)^-$ acting exclusively on the fermions. Similar to
\Ref{N2SO2} above we can conveniently incorporate these
invariances into our framework by defining three extra generators with
label $\bar\CM=\bar P=1,2,3$, satisfying 
\ba X^{\bar P}&=&0\;,\qquad [\CS^{\bar P},\CS^{\bar
Q}]~=~\varepsilon^{\bar P\bar Q\bar R}\,\CS^{\bar R}\;, 
\label{N4SO3}
\ea
so that the four-by-four matrices $\CS^{\bar P}$ generate the $\SO3^-$
group on the fermions. With these definitions, \Ref{conN4} implies 
\ba
\GTh_{\bar P\bar Q} &=& \Gth \,\delta_{\bar P\bar Q}\quad
\Longleftrightarrow\quad
T^{\bar P\bar Q}~=~\Gth \,\delta^{\bar P\bar Q}
\;.
\ea
Assuming that $\theta\not=0$, the $SO(3)^-$ gauge transformations on
the spinors can 
combine with possible target space isometries, provided that the
isometry group contains an $SO(3)$ subgroup. The gauge group
decomposes into a direct product ${\rm G}_0\times\SO3$, where the 
gauge symmetries associated with ${\rm G}_0$ correspond to the
embedding tensor,
\ba
\GTh_{\cM\cN} - \GTh_{\cM\bar P}\,\GTh^{\bar P\bar Q}\,\GTh_{\bar Q\cN}
\;,
\ea
with $\GTh^{\bar P\bar Q}$ the inverse of $\GTh_{\bar P \bar Q}$. 

When $\GTh_{\bar P\bar Q}=0$, the gauge group can still include the
$\SO3^-$ R-symmetry group (or an $\SO2$ subgroup thereof) through the mixed
components of the embedding tensor. Hence we distinguish three
$\SO3$ generators labeled by $P=1,2,3$, 
\ba
\label{SO3-KV}
T_P &=& \GTh_{P\cM} X^\cM + \GTh_{P\bar P} \,\CS^{\bar P}\;,
\ea
where  $\GTh_{P\cM} X^\cM$ denote possible corresponding $\SO3$ Killing
vectors. The
presence of the mixed components $\GTh_{P\bar P}$ induces another set
of gauge group generators,
\ba
T_{\bar P} &=& \GTh_{\bar P Q}\,X^Q \;, 
\ea
where the $\GTh_{\bar P Q}X^Q$ denote three more Killing
vectors. From~\eqn{thetapm} it then follows that the generators
$T_{\bar P}$ are mutually commuting and transform as a vector under
the $\SO3$ isometries \eqn{SO3-KV}. The $\SO3$ Killing vectors
$\GTh_{P\cM}X^\cM$ must be nonvanishing. Hence the gauge group
consists of a semidirect product of $\SO3$ with the three-dimensional
abelian group ${\cal T}$ generated by the $T_{\bar P}$, possibly
multiplied with another subgroup of the isometry group. These are the
type of theories one obtains upon dimensional reduction of
four-dimensional $N=2$ supergravity without hypermultiplets and a
gauged $\SU2$ subgroup of the R-symmetry group \cite{dWiVPr84}. It is
possible to restrict the embedding tensor, such that we gauge only an
$\SO2$ subgroup of the $\SO3^-$ R-symmetry group. By similar reasoning
as above, it follows that one must at least have an $\SO2\times\SO2$
gauge group.

\section{Symmetric spaces}
For $N>4$, it has been shown in \cite{dWToNi93} that the target spaces
are symmetric homogeneous spaces
$G/H$ such that $d=\dim{\rm G}-\dim{\rm H}$. A list of these
spaces is given in table~\ref{GG1}.
In this section, we show that the underlying group structure
allows to translate the consistency condition for admissible gauge
groups into a projection equation for the embedding
tensor~$\GTh$. This provides an efficient way of classifying and
constructing solutions to this equation which has been applied to the
gaugings of maximal supergravity in~\cite{NicSam00,NicSam01a}.

For a homogeneous target space manifold ${\rm G}/{\rm H}$, the scalar
fields are described by means of a ${\rm G}$-valued matrix $\Mat$, on
which the rigid action of ${\rm G}$ is realized by left
multiplication, while ${\rm H}$ acts as a local symmetry by
multiplication from the right. The latter gauge freedom may be used to
eliminate the spurious degrees of freedom in $\Mat$ and obtain a coset
representative so that $\Mat=\Mat(\phi^i)$ is directly parametrized by
the $d$ scalar fields $\phi^i$. In the case at hand, the group ${\rm
H}$ is a maximal compact subgroup of G and given by $\SO{N} \times
{\rm H}^\prime$.  The generators of the group G constitute a Lie
algebra $\mathfrak{g}$, which decomposes into $\{ t^{\cM}\} =
\{X^{IJ}, X^\alpha, Y^{A} \}$. The $X^{IJ}$ generate $\SO{N}$ and the
$X^\alpha$ generate the group~${\rm H}^\prime$; together they span the
subalgebra $\mathfrak{h}$ while the remaining (noncompact) generators
$Y^A$ transform in a (not necessarily irreducible) spinor
representation of $\SO{N}$.  The relevant representations are
collected in appendix~\ref{app:decompositions}.  Expanding the
dependence of $\Mat(\phi^i)$ on $\phi^i$ defines
\ba
\Mat^{-1} \dd_i \Mat &=& \ft12\,Q_i^{IJ}\, X^{IJ} +
Q^\alpha_i\, X^{\alpha} + e_i{}^A\, Y^{A} \;. 
\label{Vi}
\ea 
We note the connections $Q_i$ (the ones associated with the $X^\alpha$
were introduced at the end of subsection~2.1). The vielbein
$e_i{}^A$ may be used to convert curved target-space indices into
flat $\SO{N}$ spinor indices. The target-space metric $g_{ij}$ and the
antisymmetric tensors $f^{IJ}_{ij}$ are realized by
\ba
g_{ij} &=& e_i{}^A\,e_j{}^B\,\delta_{AB} \;,\qquad
f^{IJ}_{ij} ~=~ - \Gamma^{IJ}_{AB}\,e_i^A\,e_j^B \;,
\label{fghomogen}
\ea 
with $\SO{N}$ $\Gamma$-matrices $\Gamma^{IJ}_{AB}$.\footnote{
  Only for $N=9$ and $N=16$ the $Y^A$ transform in an irreducible
  spinor representation of $\SO{N}$. Generically, the $Y^A$ comprise a
  reducible representation of $\SO{N}\times {\rm H}'$ ({\it
  c.f.}~appendix~\ref{app:decompositions} for a complete
  list). Correspondingly, the $\Gamma$ matrices
  $\Gamma^{I}{}_{\!A\dA}$, $\Gamma^{IJ}_{AB}$ are understood
  unambiguously as acting separately on the different subspaces and as
  identity on each ${\rm H}'$-representation factor. Moreover, they
  define what we will refer to as the conjugate spinor representation
  with associated indices $\dA$. {\it E.g.}\ for $N=10$, we have
  $Y^A={\bf 16}^++\overline{\bf 16}{}^{-}$, and the fermions transform
  in the conjugate representation $\overline{\bf 16}{}^{+}+{\bf 16}^-
  $.} 
The matter fermion fields are redefined by converting their target-space
indices into indices associated with the conjugate spinor
representation of $\SO{N}$, 
\ba 
\chi^\dA &\equiv&
{1\over N}\,e_i^A\,\Gamma^I_{A\dA}\,\chi^{i I} \;.  
\ea 
All the general formulae obtained above may be conveniently
translated, noting that the projector $\bbP$ from \Ref{projchi}
factorizes according to
\ba
g_{ik}\,\bbP_J^I{}_j^k &=& {1\over N} \left(e_i^A \GG^I_{A\dA}\right) 
\left(\GG^J_{B\dA} e_j^B\right) \;.  
\ea

The isometries are generated by the left
action of ${\rm G}$ on $L(\phi)$, accompanied by a compensating 
${\rm H}$-transformation to remain in the coset representative, 
\ba
X^{\cM\,i}\,\partial_i \Mat &=& t^\cM \Mat - \Mat\, {\cal S}^\cM(\phi^i) \;,
\qquad {\cal S}^\cM(\phi^i) \in \mathfrak{h} \;.
\label{Mgh}
\ea
Here $\CS^\cM$ decomposes into $\CS^{\cM\,IJ}$ (these quantities were
already introduced in a more general context in section~2.2) and
$\CS^{\cM\,\alpha}$, belonging to $\SO{N}$ and ${\rm H}^\prime$,
respectively. Subsequently one forms the combinations $\CV^\cM \equiv
\CS^\cM +  X^{\cM\,i}\, Q_i$ for all components belonging to
$\mathfrak{h}$. For any coset space one can 
show~\cite{dWit02} that these $\CV^\cM$, together with the
$\CV^{\cM\,i}\equiv X^{\cM\,i}$ are subject to a system of linear
first-order differential equations, which 
includes the generators of H and the curvatures of the connections
$Q_i$. For the case at hand the resulting equations coincide
precisely with \eqn{DV}. The $\CV^{\cM\,i}$ can also generally be
expressed in terms of the coset representatives $\Mat$, and the combined
expression for all the $\CV$ takes the following form,  
\ba 
\Mat^{-1} t^\cM \Mat ~\equiv~  \CV^\cM{}_{\cA}\,t^\cA ~=~ 
\ft12\,\CV^{\cM\,IJ}\, X^{IJ} +
\CV^\cM{}_{\alpha}\, X^{\alpha} + \CV^\cM{}_{A}\, Y^{A} \;. 
\label{dA}
\ea
where $\CV^{\cM\, i} = g^{ij} e_j{}^A \CV^\cM{}_A$. Hence the $\CV$
span an element in the Lie algebra $\mathfrak{g}$, which coincides
with the algebra of the generators of the isometries. At this point we
can make direct contact with the map~\Ref{Vmap}, which now defines an
isomorphism, corresponding to the field-dependent
conjugation~\Ref{dA}. In particular, the $T$-tensor~\Ref{T-tensors},
given by
\ba
T_{\cA\cB} &=& \CV^\cM{}_{\cA}\,\GTh_{\cM\cN}\,\CV^\cN{}_{\cB}
\;,
\label{Thom}
\ea
where ${\cal A}= \{IJ,\alpha,A\}$, 
contains the embedding tensor of the gauge group as
$\GTh=T|_{\CV={\rm I}}$. 

\subsection{Lifting the consistency constraints}
\label{lifting}

We recall that the consistency condition for a supersymmetric gauging
takes the form of a single equation \Ref{con88} for the $T$-tensor,
and dictates the absence of the $\SO{N}$
representation~$\Yboxdim4pt{\yng(2,2) }$ in $T^{IJ,KL}$. In
order to satisfy this equation on the entire scalar manifold ${\rm
G}/{\rm H}$, the structure \Ref{Thom} of the $T$-tensor shows that the
entire ${\rm G}$-orbit of the $\SO{N}$
representation~$\Yboxdim4pt{\yng(2,2) }$ must vanish. Consider
now the decomposition of the $T$-tensor under $\rm G$, to
\ba
\Adj \times_{\rm sym} \Adj &=& 
{\bf 1} \oplus \Big[{\textstyle{\bigoplus_{i}}}\,  R_{i}\Big]
\;,
\qquad
\Longrightarrow \quad
T_{\cA\cB} ~=~ \Gth \, \eta_{\cA\cB} +
{\textstyle{\sum_i}}\; 
\bbP\!^\vl_{R_{i}}\,T_{\cA\cB} \;,
\label{Tdec}
\ea
where ${\bf 1}$ and $\Adj$ are the trivial and the adjoint
representation of $\rm G$, respectively, $\eta_{\cA\cB}$ is the
Cartan-Killing form of $\rm G$, and ``$\times_{\rm sym}$'' denotes the
symmetrized tensor product. By $\bbP\!^\vl_{R_{i}}$ we denote the $\rm
G$-invariant projector onto the representation $R_{i}$. From the
$\SO{N}$ composition of the generators, it is clear that there is a
unique irreducible representation $R_0$ of $\rm G$ appearing in the
sum in \Ref{Tdec} that branches under $\SO{N}$ such that it contains
the representation $\Yboxdim4pt{\yng(2,2)}$. The condition \Ref{con88}
is thus equivalent to
\ba
\bbP\!^\vl_{R_{0}}\,T_{\cA\cB} &=& 0 \;.
\label{projT}
\ea
The other $\SO{N}$ representations contained in this equation can be
obtained explicitly by successively taking derivatives
of~\Ref{con88}. Because \Ref{projT} is a $\rm G$-covariant condition,
it is also equivalent to
\ba
\bbP\!^\vl_{R_{0}}\,\GTh_{\cM\cN} &=& 0 \;.
\label{projTh}
\ea

The underlying coset structure thus allows to translate the {\em
field-dependent} form \Ref{relT} of the consistency condition into a
single condition for the {\em constant} embedding tensor of the gauge
group ${\rm G}_0$. After identifying the representation $R_0$, the
condition for the embedding tensor corresponding to a consistent
gauging can thus be given in explicit form. In table~\ref{GG1}, we
have collected the decompositions \Ref{Tdec} and the representations
$R_0$ for all theories with $N>4$.\footnote{We used the LiE package
\cite{LeCoLi92} for computing the decompositions of tensor products
and the branching of representations; throughout this paper we use the
corresponding conventions for the Dynkin weights.}  Given a subgroup
${\rm G}_0\subset {\rm G}$ with a corresponding embedding tensor
$\GTh_{\cM\cN}$, equation \Ref{projTh} provides a simple and efficient
criterion for checking whether ${\rm G}_0$ can be consistently gauged
while preserving all supersymmetries. The solutions of \Ref{projTh}
will be referred to as {\em admissible gauge groups} ${\rm G}_0$.  In
the following two sections, we discuss some of these solutions, case
by case for the different~$N$. We close this section with some general
remarks on the solutions of~\Ref{projTh}.

\begin{table}[bt]
\centering
\begin{tabular}{|c|c|c||c|l|} \hline
$N$ & ${\rm G}/{\rm H}$ & $d$ &  ${\Adj}$ &
$\Adj \times_{\rm sym} \Adj$
\\ \hline\hline
5& $\frac{\Sp{2,k}}{\Sp{2}\times\Sp{k}}$ 
& $8k$ & 
 $(2,0,\dots)$& 
$
(0,\dots)+ (0,1,\dots)+(0,2,\dots)+
\underline{(4,0,\dots)}$
\\ \hline
6& $\frac{\SU{4,k}}{{\rm S}({\rm U}(k)\times{\rm U}(4))}$
& $8k$&
$(1,\dots,1)$ &
$(0,\dots,0)+(1,\dots,1)+(0,1,\dots,1,0)+
\underline{(2,\dots,2)}$
\\ \hline
8& $\frac{\SO{8,k}}{\SO8\times\SO{k}}$
& $8k$ &
$(0,1,\dots)$ 
&
$(0,0,\dots)+(0,0,0,1,\dots)+(2,0,\dots)
+\underline{(0,2,\dots)}$
\\ \hline
9& $\frac{ {\rm F}_{4(-20)}}{\SO9} $
& 16 &
${\bf 52}$ &
$ {\bf 1}+ {\bf 324}+\underline{{\bf 1053}}$
\\ \hline
10& $\frac{ {\rm E}_{6(2)}}{\SO{10}\times{\rm U}(1)} $
& 32 &
${\bf 78}$ &
$ {\bf 1}+{\bf 650}+\underline{{\bf 2430}}$
\\ \hline
12& $\frac{ {\rm E}_{7(-5)}}{\SO{12}\times\Sp1} $
& 64 &
${\bf 133}$ &
${\bf 1}+{\bf 1539} +\underline{{\bf 7371}}$
\\ \hline
16& $\frac{ {\rm E}_{8(8)}}{\SO{16}} $
& 128 &
${\bf 248}$ &
$ {\bf 1}+{\bf 3875} + \underline{{\bf 27000}}$
\\ \hline
\end{tabular}
\caption{\small Symmetric spaces for $N>4$. The representation $R_{0}$
  from \Ref{projTh} is underlined in the decomposition
  \Ref{Tdec}. Dots $'\dots'$ represent zero weights.}
\label{GG1}
\end{table}

A direct consequence of the projection equation \Ref{projTh} is that
the Cartan-Killing form of ${\rm G}$ is a solution to this equation as
it corresponds to the singlet in the decomposition of
\Ref{Tdec}. Therefore the full isometry group ${\rm G}$ is always an
admissible gauge group. The potential of the corresponding gauged
theory reduces to a cosmological constant because the dependence on
the scalars disappears as a result of the ${\rm G}$-invariance of the
potential. In fact all scalars fields may simply be gauged away by
means of the gauged isometries.  Apart from this trivial solution, one
may distinguish different classes of solutions of
\Ref{projTh}: $(i)$ compact gauge groups, of which in general there
are very few, $(ii)$ semisimple noncompact gauge groups, $(iii)$
non-semisimple gauge groups, and $(iv)$ complex gauge groups embedded
in the real group ${\rm G}$~\cite{FiNiSa03}. In the following, we
restrict the discussion to some semisimple solutions of~\Ref{projTh};
non-semisimple gauge groups may generically be obtained by boosting
the embedding tensors of their semisimple cousins,
(see~\cite{FiNiSa03} for a detailed discussion in the $N=16$ theory).

For a compact gauge group, the components $\GTh_{IJ\,A}$,
$\GTh_{IJ\,\alpha}$, and $\GTh_{AB}$ of the embedding tensor
vanish. Its $\SO{N}$ part $\GTh_{IJ,KL}$ must satisfy \Ref{relT}, {\it
i.e.}\ it must be of the form
\ba
\GTh_{IJ,KL} &=& \Gth\,\delta^{KL}_{IJ} +
\delta\undersym{_{I[K}\,\Xi_{L]J}\,} + \Xi_{IJKL} \;,
\label{conTH}
\ea
with a traceless symmetric tensor $\Xi_{IJ}$ and a completely
antisymmetric tensor $\Xi_{IJKL}$. Explicit inspection of~\Ref{projTh}
shows that for $N>5$ the embedding tensor must moreover satisfy
$\Gamma_{AB}^{IJKL}\,\Xi_{IJKL} \equiv 0$, which implies
$\Xi_{IJKL}\equiv0$, except for $N=8$ where the fourfold antisymmetric
product of vector representations becomes reducible.

Let us therefore consider in some more detail compact gauge groups
with embedding tensor of the form 
\ba
\GTh_{IJ,KL} &=& \Gth\,\delta^{KL}_{IJ} +
\delta\undersym{_{I[K}\,\Xi_{L]J}\,}  \;.
\label{THcompact}
\ea
It is straightforward to verify, that the choice
\ba
\Xi_{IJ} &=& \left\{ 
\begin{array}{rl} 2(1\mis \ft{p}{N})\, \delta_{IJ} & 
\mbox{for}\;\; I\le p \\[.5ex]
-2\ft{p}{N}\,\delta_{IJ} & \mbox{for}\;\; I>p 
\end{array} \right.
\;,\qquad
\Gth ~=~ \frac{2p-N}{N} \;,
\label{SOpq}
\ea
describes the embedding of $\SO{p}\times \SO{N\mis p}\subset \SO{N}$ as
$\GTh=\GTh^{\SO{p}} - \GTh^{\SO{N-p}}$, {\it i.e.}\ with opposite
coupling constant. This ratio is fixed by the requirement that the
embedding tensor takes the form \Ref{THcompact}. Note that $\Gth=0$
can only be achieved for even $N=2p$. Likewise, one can check that no
product $\SO{p_1}\cro \dots \cro \SO{p_n}$ with more than two factors
can be embedded into $\SO{N}$ with an embedding tensor of the form
\Ref{THcompact}. This severely restricts the possible choices of
compact gauge groups.

\mathon
\subsection{The theories with $8<N\le 16$}
\mathoff

For the theories with $N=9, 10, 12, 16$, the physical fields form a
single supermultiplet, out of which the scalars parametrize the
exceptional coset spaces $\EF{4}{-20}/\SO9$, $\EE{6}{-14}/(\SO{10}\cro
\UU{1})$, $\EE{7}{-5}/(\SO{12}\cro \SU{2})$, and $\EE{8}{8}/\SO{16}$,
respectively. The decomposition \Ref{Tdec} for all these groups
contains three irreducible representations only ({\it c.f.}\
table~\ref{GG1}), so that the embedding tensor of an admissible gauge
group \Ref{projTh} is entirely contained in the union of a single
$G$-representation $R_1$ and the singlet
\ba
\GTh_{\cM\cN} &=& \Gth\,\eta_{\cM\cN} +
\bbP\!^\vl_{R_{1}}\,\GTh_{\cM\cN} \;.
\label{admdec}
\ea
This enables one to identify solutions of \Ref{projTh} by purely
group-theoretical reasoning as we shall summarize in the following
observations.
\begin{itemize}
\item
Let ${\rm G}_0\subset {\rm G}$ be a semisimple subgroup of ${\rm G}$
such that the decomposition of $R_0$ from \Ref{projTh} under ${\rm
G}_0$ does not contain a singlet. Then ${\rm G}_0$ is an admissible
gauge group.
\item 
Let ${\rm G}_0={\rm G}^{(1)}\times {\rm G}^{(2)}\subset {\rm G}$ be a
semisimple subgroup of ${\rm G}$ such that the decomposition of~$R_0$
from \Ref{projTh} under~${\rm G}_0$ contains precisely one
singlet. Then ${\rm G}_0$ is an admissible gauge group, provided a
fixed ratio of coupling constants of ${\rm G}^{(1)}$ and ${\rm
G}^{(2)}$~\cite{NicSam01a}. We denote its embedding tensor by
\ba
\GTh_{\cM\cN} &=& 
g_1 \,\GTh^{(1)}_{\cM\cN} + g_2 \,\GTh^{(2)}_{\cM\cN} \;,
\label{g1g2}
\ea
where $\GTh^{(1)}_{\cM\cN}$ and $\GTh^{(2)}_{\cM\cN}$ are the
restrictions of the Cartan-Killing form~$\eta_{\cM\cN}$ of~${\rm G}$
onto ${\rm G}^{(1)}$ and ${\rm G}^{(2)}$, respectively.

\item
Let ${\rm G}_0={\rm G}^{(1)}\times {\rm G}^{(2)}\subset {\rm G}$ be a
semisimple subgroup of ${\rm G}$ satisfying the above assumptions with
embedding tensor~\Ref{g1g2}. Let moreover ${\rm G}_{\cal N}\subset
{\rm G}$ be a group such that the decomposition of $R_1$ in
\Ref{admdec} under ${\rm G}_{\cal N}$ contains no singlet. Then the
ratio of coupling constants in \Ref{g1g2} is given by
\ba
\frac{g_1}{g_2} &=& - \frac{\dim {\rm G}_{\cal N} \,\dim {\rm G}^{(2)} - 
\dim {\rm G}\, \dim ({\rm G}^{(2)}\cap {\rm G}_{\cal N})  }
{\dim {\rm G}_{\cal N} \,\dim {\rm G}^{(1)} - 
\dim {\rm G} \, \dim ({\rm G}^{(1)}\cap {\rm G}_{\cal N})  }
\;.
\label{GN}
\ea
This is shown by contracting \Ref{g1g2} over ${\rm G}_{\cal N}$ and over
the full group ${\rm G}$.
\end{itemize}

Using these facts, we now give a brief case by case discussion
of some of the semisimple admissible gauge groups for the theories
with $N>8$.

\mathon
{\bf $N=16\,$}:$\;\;$
\mathoff
The gaugings of the maximal three-dimensional supergravity have been
constructed and discussed in great detail in
\cite{NicSam00,NicSam01a}; we include some of the results here for
completeness. The embedding tensor of an admissible compact gauge
groups must take the form \Ref{THcompact}. However, the explicit
decomposition of the $T$-tensor \Ref{exT16} shows that the two
singlets in~\Ref{exT16} are linearly related (specifically
$8\,\GTh_{IJ,IJ}=-15\,\GTh_{AA}$). Since $\GTh_{AB}=0$ for a compact
gauge group, this requires $\Gth=0$. From \Ref{SOpq} it then follows
that the only compact admissible gauge group is the product
$\SO8\times \SO8$ with opposite gauge coupling constants. The
noncompact admissible gauge groups include the $\SO{p,8\mis p}\times
\SO{p,8\mis p}$, but also the exceptional groups $\EF{4}{-20}\cro
\EG{2}{-14}$, $\EE{6}{-14}\cro \SU{3}$, $\EE{7}{-5}\cro \SU{2}$, and
different real forms thereof (see~\cite{NicSam01a} for a detailed
list). They all satisfy the assumptions leading to~\Ref{g1g2}. The
ratios of coupling constants for these groups are straightforwardly
derived from \Ref{GN} with ${\rm G}_{\cal N}=\SO{16}$.

\mathon
{\bf $N=12\,$}:$\;\;$
\mathoff
There are three singlets in the decomposition \Ref{exT12} related by a
single linear condition. This leads to a larger variety of compact
gauge groups. Roughly speaking, a nonvanishing $\Gth$ in
\Ref{THcompact} may be compensated by switching on the extra $\SU{2}$
factor from ${\rm H}=\SO{12}\cro \SU{2}$ in the gauge group, such that
the noncompact part $\GTh_{AB}$ still remains zero. For example, the
decomposition of \Ref{Tdec} under ${\rm H}$ shows that this subgroup
itself satisfies the assumptions leading to \Ref{g1g2}; unlike in the
maximal case, ${\rm H}$ is itself an admissible gauge group with a
fixed ratio of coupling constants. This ratio can be derived from
\Ref{GN} (using for example ${\rm G}_{\cal N}=\SU{6,2}$) and gives
$\GTh=\GTh_{\SO{12}}-3\GTh_{\SU{2}}$. In general, the admissible
compact gauge groups are given by the products $\SO{p}\cro \SO{12\mis
p}\cro \SU{2}$ with embedding tensor
\ba
\GTh &=& \GTh_{\SO{p}} - \GTh_{\SO{12-p}}
+\ft12(6\mis p) \, \GTh_{\SU{2}} \;.
\label{THex12}
\ea
The relative coupling constant between the two first factors is
determined by \Ref{SOpq}, while the relative factor in front of the
last term stems from the relation $3\GTh_{IJ,IJ}=
-22\,\GTh_{\Ga\Ga}$ for compact gauge groups ({\it c.f.}~\Ref{exT12})
whose relative coefficient may be fixed from the case $p=12$ given
above. Note that (only) for $p=6$, the gauge group lies entirely in
$\SO{12}$ and the $\SU{2}$ factor is not gauged. Among the
noncompact admissible gauge groups there are $\EE{6}{2}\cro \UU{1}$,
$\EF{4}{-20}\cro \SU{2}$, $\EG{2}{2}\cro \Sp{3}$, all of which are
maximal subgroups of $\EE{7}{-5}$.

\mathon
{\bf $N=10\,$}:$\;\;$
\mathoff
Similar to the above, the admissible compact gauge groups in
this case are the products $\SO{p}\cro \SO{10\mis p}\cro \UU{1}$ with
embedding tensor
\ba
\GTh &=& \GTh_{\SO{p}} - \GTh_{\SO{10-p}}
+\ft13(5\mis p) \,  \GTh_{\UU{1}}  \;.
\label{THex10}
\ea
The relative coupling constants are fixed as in~\Ref{THex12},
using for example ${\rm G}_{\cal N}=\Sp{2,2}$. For $p=5$, the gauge group
lies entirely in the $\SO{10}$ and the $\UU1$ factor is not
gauged. Among the noncompact admissible gauge groups there are
$\SU{4,2}\cro \SU2$, $\EG{2}{-14}\times \SU{2,1}$, as well as the simple
group $\EF{4}{-20}$. All these gauge groups are maximal
subgroups in $\EE{6}{-14}$.

\mathon
{\bf $N=9\,$}:$\;\;$
\mathoff
In this case there is no additional factor in ${\rm H}$; however, the
explicit decomposition \Ref{exT9} shows that the two singlets
appearing are independent. Therefore, again a compact gauge group does
not necessarily require vanishing $\Gth$ in \Ref{THcompact}. In
particular, the group ${\rm H}=\SO9$ itself is an admissible gauge
group. More generally, the admissible compact gauge groups are the
products $\SO{p}\cro \SO{9\mis p}$ with embedding tensor
\ba
\GTh &=& \GTh_{\SO{p}} - \GTh_{\SO{9-p}}\;.
\ea
Among the noncompact admissible gauge groups there are
$\EG{2}{-14}\cro \SL2$ and $\Sp{2,1}\cro \SU2$ which are maximal
subgroups of $\EF{4}{-20}$.

\mathon
\subsection{The theories with $4<N\le 8$}
\mathoff
For $N=5, 6, 8$, the field content of the ungauged theories is given
by an arbitrary number $k$ of supermultiplets whose scalars fields
parametrize the coset spaces $\Sp{2,k}/(\Sp2\cro
\Sp{k})$, $\SU{4,k}/{\rm S}(\UU2\cro \UU{k})$, and $\SO{8,k}/(\SO8\cro
\SO{k})$, respectively.

\mathon
{\bf $N=8\,$}:$\;\;$
\mathoff
The $N=8$ ungauged theories have been constructed in \cite{MarSch83},
their gaugings were discussed in~\cite{NicSam01b}.\footnote{Note, that
our conventions here differ from those used in
\cite{MarSch83,NicSam01b} by a triality rotation ${\bf
8_v}\leftrightarrow{\bf 8_s}$ of $\SO8$ in order to fit into the
general scheme.} Consistency of the gauging is again encoded in
\Ref{projTh} with $R_0=(0, 2, 0, 0, \dots)$.\footnote{In
\cite{NicSam01b}, a slightly stronger consistency condition had been
given, namely simultaneous absence of the $(2, 0, 0, 0, \dots)$. This
is in general not necessary. Restricting to compact gauge groups
$G_0\subset \SO8$ however, the two conditions turn out to be
equivalent.}  Just as in the previous examples, one may consistently
gauge the entire compact subgroup ${\rm H}=\SO8\times \SO{k}$ with a
fixed ratio between the two coupling constants. More interesting are
the admissible gauge groups that lie entirely in the
group~$\SO8$. Note that for $N=8$, the condition
$\Gamma^{IJKL}_{AB}\GTh_{IJ,KL}=0$, no longer forces the entire
antisymmetric part $\GTh_{[IJ,KL]}$ of the embedding tensor to vanish,
rather the embedding tensor of a compact gauge group ${\rm G}_0\subset
\SO8$ takes the general form
\ba
\GTh_{IJ,KL} &=&
\delta\undersym{_{I[K}\,\Xi_{L]J}\,} + \Xi_{IJKL} \;,
\label{THcompact8}
\ea
with a traceless symmetric tensor $\Xi_{IJ}$ and selfdual
$\Xi_{IJKL}=\frac1{24}\Geps_{IJKLPQRS}\,\Xi_{PQRS}$, relaxing
\Ref{THcompact}. In the standard way \Ref{SOpq}, the group $\SO4\times
\SO4$ may be embedded with relative coupling constant of $-1$. A
nonvanishing $\Xi_{IJKL}$ in \Ref{THcompact8} furthermore allows
to introduce an arbitrary relative coupling constant $\alpha$ between
the two factors inside of each $\SO4$ (see~\cite{NicSam01b} for
details). This corresponds to the existence of the one-parameter family
${\rm D}^1(2,1;\alpha)$ of $N=4$ superextensions of the AdS group
$\SO{2,2}$, which appear as spacetime isometries.

\mathon
{\bf $N=6\,$}:$\;\;$
\mathoff
Closer inspection of the decomposition \Ref{exT6} of $\Theta$ shows
that the maximal compact subgroup ${\rm H}=\SU4\cro \SU{k}\cro \UU1$
is among the admissible gauge groups with an embedding tensor which
forms a linear combination of the corresponding singlets in
\Ref{exT6}. Since a compact gauge group requires $\GTh_{AB}=0$, and
the four singlets appearing in \Ref{exT6} are linearly related, there
are only two independent coupling constants. Other compact gauge
groups are obtained by replacing the $\SU4$ factor by one of its
subgroups $\SO{p}\cro \SO{6\mis p}$, the embedding tensor taking the
form~\Ref{SOpq}. The total embedding tensor is given by
\ba
\GTh &=& \GTh_{\SO{p}} - \GTh_{\SO{6-p}} +  \alpha\, \GTh_{\SU{k}} -
\frac{4\alpha(k-1)+ k(p-3)}{4+k}\,\GTh_{\UU1} \;,
\label{ratio6}
\ea
with a free parameter $\alpha$. The relative coefficients are obtained
in a similar way as in \Ref{THex12}, \Ref{THex10},
generalizing~\Ref{GN} to products of more factors and using ${\rm
G}_{\cal N}=\SO{4,k}$. The only compact admissible gauge group which
lies entirely in~$\SO6$ is its subgroup~$\SO3\cro \SO3$.

\mathon
{\bf $N=5\,$}:$\;\;$
\mathoff
The explicit decomposition of the $T$-tensor \Ref{exT5} shows that the
group $\Sp2\sim \SO5$ as well as its product with the entire $\Sp{k}$
and independent coupling constants, are admissible gauge groups. The
embedding tensors are given by a linear combination of the two
corresponding singlets in the decomposition \Ref{exT5}. Instead of
$\Sp2$ one may also gauge any of its subgroups $\SO4$, or $\SO2\cro
\SO3$, the embedding tensor taking the form~\Ref{SOpq}. From
\Ref{exT5} it follows moreover that for $N=5$ even the completely
antisymmetrized tensor $\GTh_{[IJ,KL]}$ may be nonvanishing for a
compact gauge group, since the two $({\bf 5},{0,0,0,\dots})$
representations in \Ref{exT5} may be chosen independently.

\mathon
\subsection{$N=4$: the symmetric spaces}
\mathoff
\label{N4symmetric}

Recall, that for $N=4$ the target space manifold is a product of two
quaternionic spaces and consistency of the gauged theory is encoded in
equation~\Ref{conN4} for the $T$-tensor to be satisfied on the entire
target space manifold. In case, the quaternionic spaces are symmetric,
one can apply the techniques described in this section to lift
equation~\Ref{conN4} to an algebraic projection equation on the
embedding tensor, and exploit the known decomposition of the isometry
group under $\SO4$ to perform a similar analysis of the admissible
gauge groups. To this end, we list all quaternionic symmetric
spaces~\cite{Wolf65} together with the decompositions~\Ref{Tdec} and
the representations $R_0$ defining~\Ref{projTh} in table~\ref{GG2}. We
leave the further study of these gaugings and the admissible gauge
groups to the reader.

\begin{table}[bt]
\centering
\begin{tabular}{|c|c|c|l|} \hline
${\rm G}/{\rm H}$ & $d$ &  ${\Adj}$ &
$\Adj \times_{\rm sym} \Adj$
\\ \hline\hline
$\frac{\Sp{m,1}}{\Sp{m}\times\Sp1}$ 
& $4m$ & $(2,0,\dots)$ &
 $(0,0,\dots)+ (0,1,\dots)+(0,2,\dots)+
\underline{(4,0,\dots)}$
\\ \hline
$\frac{\SU{m,2}}{{\rm S}({\rm U}(m)\times{\rm U}(2))}$
& $4m$& $(1,\dots,1)$ &
$(0,\dots,0)+(0,1,\dots,1,0)+(1,\dots,1)+
\underline{(2,\dots,2)}$
\\ \hline
$\frac{\SO{m,4}}{\SO{m}\times\SO4}$
& $4m$& $(0,1,\dots)$ &
$(0,0,\dots)+(0,0,0,1,\dots)+(2,0,\dots)+\underline{(0,2,\dots)}$
\\ \hline
$\frac{ {\rm G}_{2(2)}}{\SO4} $
& 8 & ${\bf 14}$ &
$ {\bf 1}+ {\bf 27} +\underline{{\bf 77}}$
\\ \hline
$\frac{ {\rm F}_{4(4)}}{\Sp3\times\Sp1} $
& 28 &${\bf 52}$ &
$ {\bf 1}+ {\bf 324}+\underline{{\bf 1053}}$
\\ \hline
$\frac{ {\rm E}_{6(2)}}{\SU6\times\Sp1} $
& 40 &${\bf 78}$ &
$ {\bf 1}+{\bf 650}+\underline{{\bf 2430}}$
\\ \hline
$\frac{ {\rm E}_{7(-5)}}{\SO{12}\times\Sp1} $
& 64 &${\bf 133}$ &
${\bf 1}+{\bf 1539} +\underline{{\bf 7371}}$
\\ \hline
$\frac{ {\rm E}_{8(-24)}}{{\rm E}_7\times \Sp1} $
& 112 &${\bf 248}$ &
$ {\bf 1}+{\bf 3875} + \underline{{\bf 27000}}$
\\ \hline
\end{tabular}
\caption{\small Symmetric spaces for $N=4$. The representation $R_0$
  from \Ref{projTh} is underlined.}  
\label{GG2}
\end{table}

\section{Concluding remarks}
We have constructed the general $N$-extended gauged supergravity
theories in three space-time dimensions. The gaugings constitute
supersymmetric deformations of the ungauged theories of
\cite{dWToNi93}, and are entirely characterized by a symmetric
embedding tensor $\GTh_{\cM\cN}$ that specifies the gauge group as a
subgroup of the full invariance group 
of the ungauged theories. This invariance group consists of the
target-space isometries, and (for $N=2,4$) possible R-symmetry
transformations. The embedding tensor must generate a proper
subgroup and must be invariant under the gauge group. This is
expressed by the conditions~\Ref{subgrouptheta}. Supersymmetry imposes
additional conditions on the embedding tensor, which are expressed in
terms of constraints on the so-called $T$-tensor. For $N>2$ these
constraints are encoded in~\Ref{relT}. We have
analyzed these constraints for the different values of $N$. Any
subgroup can be gauged for $N\leq3$. For $N=1,2$ there exist
supersymmetric deformations that are not induced by a 
gauging; for $N=2$ their presence may form an 
obstacle to certain gauge groups. For $N\geq4$ there are restrictions on the
embedding tensor and thus on the corresponding subgroups that can be
gauged. For $N>4$ all target spaces are symmetric. For symmetric
target spaces with $N\geq4$ the restriction on the gauge group can
conveniently be formulated in terms of the algebraic projection
equation~\Ref{projTh} on the embedding tensor $\GTh$. 

The gaugings require the usual masslike terms and a scalar potential
in the Lagrangian, parametrized by three tensors $A_1$, $A_2$ and
$A_3$,
\ba
\label{gauged-L}
\CL &= &\CL_0 
+ e g \: \Big\{
\ft12  A_1^{IJ}\,\Bpsi{}^I_\mu\,\Gg^{\mu\nu}\,\psi^J_\nu\, +
 A^{IJ}_{2\,j}\,\Bpsi{}^I_\mu\,\Gg^\mu \chi^{j J} +
\ft12  A_{3\,}{}_{ij}^{IJ}\, \Bchi^{i I}\chi^{j J}\Big\} \non
&& 
+{4\, e g^2\over N}  \left( A_1^{IJ}A_1^{IJ} -\ft12 N\,
g^{ij}\, A_{2i}^{I\,J} A_{2j}^{I\,J} \right)\;,
\ea
where $\CL_0$ denotes the ungauged Lagrangian \eqn{L} with the spacetime
derivatives extended by extra covariantizations
associated with the gauging, as specified in \eqn{covscalar},
\eqn{covfermions} and \eqn{covepsilon}.
The supersymmetry transformations of the fermion fields acquire extra
terms proportional to $A_1$ and $A_2$. They read 
\begin{eqnarray}
\Gd \psi^I_\mu &=& {\cal D}_\mu \epsilon^I   - \ft18 g_{ij} \,\Bchi{}^{iI}
\gamma^\nu \chi^{jJ}\, \gamma_{\mu\nu} \,\epsilon^J - 
\Gd \phi^i \,Q_i^{IJ} \psi^J_\mu + g\,A_1^{IJ} \Gg_\mu\,\Ge^J \;,\non[.5ex]
\Gd \chi^{iI} &=& \ft12 \left(\delta^{IJ}{\bf 1} \mis f^{IJ}
\right)^i{}_{\!j}\; 
\FMslash{\widehat {\cal D}} \phi^j \, \epsilon^J -
\Gd \phi^j \left(
\Gamma^i_{jk}\,\chi^{kI} 
+ Q_j^{IJ}\chi^{iJ} \right)  -  g N \, A_{2}^{JiI} \,\Ge^J 
\;.
\label{gsusytraf}
\end{eqnarray}
The transformation rules for $e_\mu{}^a$ and $\phi^i$ remain as given
in \eqn{susytraf}, for the vector fields $A_\mu^\cM$ they have
been given in~\Ref{susyA}.

For $N>2$, the tensors $A_1$, $A_2$ and $A_3$ are uniquely given in
terms of the $T$-tensor ({\it c.f.}~\eqn{T-tensors}) by means of
\Ref{id1} and \Ref{A1T},
\ba
A_1^{IJ} &=& -\frac4{N\mis2}\,T^{IM,JM} +
\frac{2}{(N\mis1)(N\mis2)}\,\delta^{IJ}\,T^{MN,MN} \;,
\non
A_{2\,j}^{IJ} &=& 
\frac2{N}\,T^{IJ}{}_j + \frac4{N(N\mis2)}\,f^{M(I}{}_{\! j}{}^m
\,T^{J)M}{}_m + 
\frac{2\,\delta^{IJ}}{N(N\mis1)(N\mis2)}\, f^{KL}{}_j{}^m\, T^{KL}{}_m \;,
\non[.5ex]
A_{3\,}{}^{IJ}_{ij}
&=& {1\over N^2} \Big\{ -2\,D_{(i}D_{j)}A_1^{IJ} + g_{ij}\,A_1^{IJ} + 
A_1^{K[I} \,f_{ij}^{J]K} \non 
&& \hspace{9mm} 
+2\, T_{ij} \, 
\delta^{IJ}  
- 4\, D_{[i} T^{IJ}{}_{j]}-   2\, T_{k[i} \,f^{IJk}{}_{j]}  
\Big\}  \;.
\label{A123}
\ea
The cases $N=1, 2$ require a separate analysis, which was
presented in section~4. For symmetric spaces with $N\ge4$, these
results simplify considerably.

All gauged theories exhibit a potential \Ref{potential} for the scalar
fields $\phi^i$. In certain cases this potential is constant and
simply constitutes a cosmological term. In applications one is
often interested in stationary points of this potential which give
rise to anti-de Sitter or Minkowski solutions with residual
supersymmetries, or in de Sitter solutions.  Extremal points in the
maximal $N=16$ theory have been analyzed in some detail
in~\cite{NicSam01a,FiNiSa02,Fisc03}. Here we give a
generalization to arbitrary $N$ of some
essential formulae that may enable the reader to carry out a similar
analysis for the theories presented in this paper.

Stationary points of the scalar potential are characterized by
\Ref{varpotential}, which together with the second equation of
\eqn{id3} implies the following relation at the stationary
point,\footnote{In the remainder of this section the tensors $A_1$,
$A_2$ and $A_3$ are constant and equal to their values at the
stationary point.}
\ba
3\,A_1^{IK}\,A_{2j}^{K\,J} 
+ N\,g^{kl}\,A_{2k}^{IK}\,A_3{}^{KJ}_{lj}&=& 0\;.
\label{quadratic-stat}
\ea
The residual supersymmetry of the corresponding solution (assuming
maximally symmetric spacetimes) is parametrized by spinors
$\epsilon^I$ satisfying the condition,
\ba
A_{2i}^{J\,I}\,\epsilon^J =0\;, 
\label{susy-stat1}
\ea
which ensures that the fermions $\chi^{iI}$ remain invariant in a
bosonic background. Full unbroken supersymmetry implies that $A_2$
vanishes at the stationary point. From the gravitino 
variations one derives the condition, 
\ba
A_1^{IK}\,A_1^{KJ} \,\epsilon^J =  - {V_0\over 4\, g^2}\, \epsilon^I = 
{1\over N}  \left( A_1^{IJ}A_1^{IJ} -\ft12 N\,
g^{ij}\, A_{2i}^{I\,J} A_{2j}^{I\,J} \right)\,\epsilon^I\;,
\label{susy-stat2}
\ea 
where $V_0$ is the potential taken at the stationary point. Let us
emphasize that the two conditions \Ref{susy-stat1} and
\Ref{susy-stat2} are in fact equivalent by virtue of the first
equation of \eqn{id3}, so that the condition \eqn{susy-stat2} suffices
for establishing the (residual) supersymmetry. Obviously the potential
must be non-positive at the stationary point, so that the maximally
symmetric spacetime must be a Minkowski or an anti-de Sitter
spacetime.

{}From the above results it follows that residual supersymmetries are
associated with eigenvalues of $A_1^{IJ}$ equal to
$\pm\sqrt{-V_0/4\,g^2}$. The massive gravitini that may arise are
associated with different eigenvalues and span an orthogonal
subspace. We will distinguish the indices associated to this
orthogonal subspace by $\hat I,\hat J,\ldots$. The massive gravitini
can be identified by the linear combinations,
\ba
\psi^{\hat I}_{\mu\,\rm massive} \propto  \left( A_1^{\hat I\hat
K}A_1^{\hat K\hat J} +{V_0\,
\delta^{\hat I\hat J}\over 4\,g^2}\right) \psi^{\hat J}_\mu +{1\over
  2\,g} 
\,
A_{2\,j}^{\hat IJ}\,\partial_\mu \chi^{jJ}  
+\ft12 A_1^{\hat I\hat K}\,A_{2\,j}^{\hat KJ}\,\gamma_\mu\chi^{jJ}\;,
\ea
which are explicitly restricted to the orthogonal subspace. Imposing
the unitary gauge by the condition
\ba
A_{2\,j}^{\hat IJ}\,\chi^{jJ}=0\;,
\label{unitary-g}
\ea 
we can simply determine the fermionic mass matrices by projection,
\ba
\CM_{\rm gravitini} &=& g\,A_1^{IJ}\;,\non
\CM_{\rm fermions}  &=& g\,A_3{}^{IJ}_{ij} + 6 \,g \,
A_{2\,i}^{\hat KI}\left[\frac{g^2\, A_1}{4 g^2\,A_1^2 -{\bf 1} \,V_0}
\right]_{\hat K\hat L} A_{2\,j}^{\hat LJ}
\;.
\label{Mfermion}
\ea
Observe that the restriction to the indices associated with the
orthogonal subspace is crucial as 
otherwise the second mass matrix would diverge. To show that this
matrix is orthogonal to the condition \eqn{unitary-g}, one makes use of
the identity \eqn{quadratic-stat}. 

It remains to evaluate the mass matrices for the bosons, which can
simply be read off from the Lagrangian and are equal to,
\ba
\CM_{\rm vectors} &=& g^2\, \GTh_{\cM\cK}\CV^{\cK i}\CV^{\cL}{}_i
\GTh_{\cL\cN}\;,
\non
\CM^2_{\rm scalars} &=& g^2\, D_i \partial_j V\;.
\label{Mboson}
\ea
The derivatives of the potential may be explicitly evaluated using the
various identities derived previously. Moreover, the form of the
vector mass matrix together with the corresponding kinetic
Chern-Simons term shows that the physical vector masses are encoded in
the matrix $gT_{ij}=g\GTh_{\cM\cN}\CV^{\cM}{}_i\CV^\cN{}_j$, so that
eventually all mass matrices can be expressed in terms of the
$T$-tensors. This is a common feature of all gauged supergravity
theories. In case of residual supersymmetry the spectrum
\Ref{Mfermion}, \Ref{Mboson} decomposes into representations of the
appropriate superextension of the AdS$_3$ isometry group
$\SO{2,2}=\SL{2,\mathbb{R}}\times\SL{2,\mathbb{R}}$.

\bigskip
\paragraph{Acknowledgements:}

We thank M.~Haack, H.~Nicolai, M.~Trigiante and S.~Vandoren for
invaluable discussions. This work is supported in part by the EC
contracts HPRN-CT-2000-00122 and HPRN-CT-2000-00131.

\begin{appendix}

\mathon
\section{About $\SO{N}$ projectors}
\mathoff
\label{SONprojectors}

For a symmetric target space, the consistency conditions on the
$T$-tensor combine into the ${\rm G}$-invariant form \Ref{projT}. In
the general case, these consistency conditions take an $\SO{N}$
covariant form, {\it c.f.}~\Ref{T3-id}, \Ref{con0}, etc. Whereas the
$\SO{N}$ representation content of \Ref{con0} is obvious ({\it
c.f.}~\Ref{con88}), it is complicated to disentangle the independent
parts of the consistency conditions \Ref{T3-id} for the
components~$T^{IJ}{}_i$, derived from \Ref{proj-constraints},
\Ref{id1}. In this appendix, we present a systematic approach to
handle these equations by means of $\SO{N}$ covariant projectors.

The tensors $f^{IJ}$ from \Ref{fQ} define a
$d$-dimensional representation of $\SO{N}$. Consider the space of
tensors $\Xi_{IJ\,j}$ of dimension $d N^2$. Using the $f^{IJ}$, we can
build the following $\SO{N}$ covariant maps on this space
\ba
{\rm Id}_{IJ\,j}^{KL\,n} &\equiv& 
\delta_I^K\delta_J^L\delta_j^n \;, 
\qquad 
P_{IJ\,j}^{KL\,n} ~\equiv~ 
\ft1{N}\,\delta_I^K
\left(\delta_J^L\delta_j^n - f^{JL}{}_j{}^n \right) \;,\non[1ex]
\bbT_{IJ\,j}^{KL\,n} &\equiv& 
\delta_I^L\delta_J^K\delta_j^n \;, 
\qquad
P_0\,{}_{IJ\,j}^{KL\,n} ~\equiv~ 
\ft1{N} \delta_{IJ}\delta_{KL} \delta_j^n 
\;.
\label{basicO}
\ea
They satisfy the following set of relations
\ba
\bbT^2 &=& {\rm Id} \;,\qquad P^2 ~=~ P \;,\qquad 
P_0^2 ~=~ \bbT P_0 ~=~ P_0\bbT ~=~ P_0 \;,\non
P_0P\bbT  &=& \ft2{N}\,P_0 - P_0P \;,\qquad
\bbT PP_0 ~=~ \ft2{N}\,P_0 - PP_0 \;,\qquad P_0PP_0 ~=~ \ft1{N}\,P_0
\;, \non
P\bbT P &=& \ft2{N}\,P - N\,PP_0P \;,
\label{relM}
\ea
which follow from~\Ref{f-properties}. An equivalent form of the last
relation is
\ba
\bbP_{Ii}^{Km}\,\bbP_{Lm}^{Jj} +
\bbP_{Ii}^{Lm}\,\bbP_{Km}^{Jj} ~=~ 
\frac2{N}\,\delta^{KL}\,\bbP_{Ii}^{Jj} \;,
\ea
which proves to be useful in checking the absence of several cubic
fermion terms in the supersymmetry variation of the Lagrangian.

For a tensor $\Xi_{IJ\,j}$, consider the following inhomogeneous system
of linear equations
\ba
P\,\Xi &=& \Xi \;,
\qquad
({\rm Id}-\bbT )\,\Xi ~=~ 2\,Z \;. 
\label{syst1}
\ea
This system admits the unique solution $\Xi = \CT Z$ if and only if
$Z$ satisfies the projection equation
\ba
Z&=& \CP\,Z \;,
\label{projZ}
\ea
where $\CT$ and the projector $\CP$ are given by 
\ba
\CP &\equiv& 
\frac{N\, (P-P\bbT-\bbT P+\bbT P\bbT ) }{2(N\mis2)}
- \frac{N\, (P_0 -N(PP_0+P_0P)+N^2 PP_0P)}{(N\mis1)(N\mis2)}
\;,
\non
\CT &\equiv& 
\frac{N}{N\mis2}\, (P-P\bbT) + 
\frac{N^2}{(N\mis1)(N\mis2)}\, (PP_0 -N PP_0P) 
\;,
\nn
\ea
with $\CP\CP=\CP$, and $\tr \CP = d N$. The necessity of
the projection condition \Ref{projZ} follows from $2Z=({\rm
Id}-\bbT )\,P\,\Xi$ and from using the relation
\ba
({\rm Id}-\bbT )\,P &=& \CP\,({\rm Id}-\bbT )\,P \;,
\ea
which follows straightforwardly from~\Ref{relM}. Likewise, one may
verify the relations 
\ba
({\rm Id}-\bbT )\,\CT &=& 2\CP \;,\qquad 
P\,\CT ~=~ \CT ~=~ \CT\,\CP\;,
\ea
which ensure that $\Xi = \CT Z$ together with \Ref{projZ} solves
\Ref{syst1}. 

This algebra can be applied to perform a systematic analysis of the
constraint equations in section~\ref{section:gauging}. As an example,
note that the antisymmetric part of the first equation in \Ref{id1}
together with \Ref{proj-constraints} constitutes a system of type
\Ref{syst1} with $Z=\ft2{N}\,T$. Its solubility thus implies the
consistency relation
\ba
T &=& \CP T \;,
\label{conT2}
\ea 
on $T^{IJ}{}_i$ which precisely agrees with \Ref{T3-id}. Since $\CP$
is a projector, this shows that equation~\Ref{T3-id} indeed describes
a closed set of consistency relations with nontrivial solution. The
tensor $A_2$ is given as $A_2 = \ft2{N}\,\CT T$ which agrees with
\Ref{id1} upon eliminating $A_1$ by means of \Ref{A1T}. The proof
again makes use of the constraint \Ref{conT2} on $T^{IJ}{}_i$.

\mathon
\section{Explicit decompositions of the $T$-tensor}
\mathoff

\label{app:decompositions}

The representation content of the $T$-tensor under the group ${\rm G}$
for the various values of $N>4$ has been given in table~\ref{GG1}. In
this appendix, we give the explicit decomposition of the $T$-tensor
under the compact group ${\rm H}=\SO{N}\cro {\rm H}'$. Since the
embedding tensor of the gauge group is obtained as $\GTh = T_{\CV={\rm
I}}$, it satisfies the same decomposition. This has been used in the
main text to further analyze the admissible compact gauge groups in
sections~5.2 and~5.3.

\bigskip

\mathon
{\bf $N=16\,$}:$\;\;$
\mathoff
Under $\SO{16}$, the adjoint representation of $\EE{8}{8}$ decomposes into
\ben
X^{IJ}:~{\bf 120}\;, \qquad Y^A: ~ {\bf 128}\;,
\een
implying that the $T$-tensor
of the gauged theory consists of
\ba
T^{IJ,KL} &=&  
{\bf 1} +{\bf 135} + {\bf 1820} 
\;,
\non
T^{AB} &=& 
{\bf 1} +{\bf 1820} 
\;,
\non
T^{IJ,A} &=& {\bf \overline{1920}}
\;,
\label{exT16}
\ea
where the two singlets ${\bf 1}$ and the two representations ${\bf
1820}$ coincide. 

\bigskip

\mathon
{\bf $N=12\,$}:$\;\;$
\mathoff
Under $\SO{12}\times \SU{2}$, the adjoint representation
of $\EE{7}{-5}$ decomposes into 
\ben
X^{IJ}: {\bf (66,1)}\;,\qquad 
X^{\Ga}: {\bf (1,3)}\;,\qquad 
Y^A: {\bf (32,2)}\;,
\een
implying that the $T$-tensor consists of
\ba
T^{IJ,KL} &=&  {\bf (1,1)} + {\bf (77,1)}+ {\bf (495,1)}
\;,\quad
T^{\Ga\Gb} ~=~  {\bf (1,1)}
\;,\quad
T^{IJ\Ga} ~=~   {\bf (66,3)}\;,
\non
T^{AB} &=& {\bf (1,1)} + {\bf (495,1)} + {\bf (66,3)}
\;,
\non
T^{IJ,A} &=& {\bf (32,2)} + {\bf (352,2)}
\;,
\non
T^{\Ga A} &=& {\bf (32,2)}
\;,
\label{exT12}
\ea
with a linear relation between the three singlets ${\bf (1,1)}$, and
where the two representations in the ${\bf (66,3)}$, ${\bf (32,2)}$,
and the ${\bf (495,1)}$, respectively, coincide.

\bigskip

\mathon
{\bf $N=10\,$}:$\;\;$
\mathoff
Under $\SO{10}\times \UU{1}$, the adjoint representation
of $\EE{6}{-14}$ decomposes into 
\ben
X^{IJ}: {\bf 45}^0\;,\qquad 
X^{\Ga}: {\bf 1}^0\;,\qquad 
Y^A:{\bf 16}^+ + {\bf\overline{16}}{}^-\;,
\een
implying that the $T$-tensor consists of
\ba
T^{IJ,KL} &=&  {\bf 1}^0 + {\bf 54}^0+ {\bf 210}^0
\;,\qquad
T^{\Ga\Gb} ~=~  {\bf 1}^0
\;,\qquad
T^{IJ\Ga} ~=~ {\bf 45}^0 \;,
\non
T^{AB} &=& {\bf 1}^0 + {\bf 10}^{+2} + {\bf 10}^{-2} 
+  {\bf 45}^0 + {\bf 210}^0 
\;,
\non
T^{IJ,A} &=& {\bf 16}^+ + {\bf \overline{16}}{}^- + 
{\bf \overline{144}}{}^+ + {\bf 144}^-
\;,
\non
T^{\Ga A} &=& {\bf 16}^+ + {\bf\overline{16}}{}^-
\;,
\label{exT10}
\ea
with a linear relation between the three singlets ${\bf 1}^0$, and
where the two representations in the ${\bf 16}^+$, ${\bf
\overline{16}}{}^-$, ${\bf 45}^0$, and ${\bf 210}^0$, respectively,
coincide.

\bigskip

\mathon
{\bf $N=9\,$}:$\;\;$
\mathoff
Under $\SO9$, the adjoint representation of $\EF{4}{-20}$ decomposes
into 
\ben
X^{IJ}:{\bf 36}\;, \qquad Y^A: {\bf 16} \;,
\een
implying that the $T$-tensor consists of
\ba
T^{IJ,KL} &=&  
{\bf 1} +{\bf 44} +{\bf 126}
\;,
\non
T^{AB} &=& 
{\bf 1} +{\bf 9} +{\bf 126}
\;,
\non
T^{IJ,A} &=& 
{\bf 16}  +{\bf 128} 
\;,
\label{exT9}
\ea
where the two ${\bf 126}$ representations coincide.

\bigskip

\mathon
{\bf $N=8\,$}:$\;\;$
\mathoff
Under $\SO8\times \SO{k}$, the adjoint representation
of $\SO{8,k}$ decomposes into
\ba
X^{IJ}: && \left(  {\bf 28} \,,\; (0, 0, 0, 0, \dots) \right)\;,
\qquad
X^{\Ga}: ~~ \left(  {\bf 1} \,,\;  (0, 1, 0, 0, \dots)\right) \;,
\non
Y^A: && \left(  {\bf 8}_s \,,\;(1, 0, 0, 0, \dots ) \right)\;,
\nn
\ea
implying that the $T$-tensor consists of
\ba
T^{IJ,KL} &=& \left( 
{\bf 1} +{\bf 35}_v +{\bf 35}_s +{\bf 35}_c \,,\; (0, 0, 0, 0, \dots) 
\;\right) 
\non
T^{\Ga\Gb} &=& \left(  {\bf 1} \,,\; (0, 0, 0, 0, \dots)
+  (2, 0, 0, 0, \dots) +  (0, 0, 0, 1, \dots) \right)
\;,
\non
T^{IJ\Ga} &=& \left(  {\bf 28} \,,\; (0, 1, 0, 0, \dots) \right)
\non
T^{AB} &=& \left(  {\bf 1} \,,\;
(0, 0, 0, 0, \dots) +  (2, 0, 0, 0, \dots) \right)
+ \left(  {\bf 35}_s \,,\; (0, 0, 0, 0, \dots) \right)
\non
&&{}
+ \left(  {\bf 28} \,,\; (0, 1, 0, 0, \dots) \right)
\;,
\non
T^{IJ,A} &=& \left(  {\bf 8}_s + {\bf 56}_s\,,\; 
(1, 0, 0, 0, \dots) \right)
\;,
\non
T^{\Ga A} &=& \left(  {\bf 8}_s \,,\; 
(1, 0, 0, 0, \dots)+(0, 0, 1, 0, \dots) \right)
\;,
\label{exT8}
\ea
with a linear relation between the three singlets $\left( {\bf 1}
\,,\; (0, 0, 0, 0, \dots) \;\right)$, and where the two
representations in the $\left( {\bf 1} \,,\; (2, 0, 0, 0, \dots)
\right)$, $\left( {\bf 35}_s \,,\; (0, 0, 0, 0, \dots) \right)$,
$\left( {\bf 8}_s\,,\; (1, 0, 0, 0, \dots) \right)$, and $\left( {\bf
28}\,,\; (0, 1, 0, 0, \dots) \right)$, respectively, coincide.

\bigskip

\mathon
{\bf $N=6\,$}:$\;\;$
\mathoff
Under $\SU{4}\times \SU{k}$, the adjoint representation
of $\SU{4,k}$ decomposes into
\ba
X^{IJ}: && \left(  {\bf 15} \,,\; (0, 0, \dots, 0, 0) \right)\;,
\qquad
X^{\Ga}: ~~ \left(  {\bf 1} \,,\;  (0, 0, \dots, 0, 0) + 
(1, 0, \dots, 0, 1)  \right) \;,
\non
Y^A: && \left(  {\bf \overline{4}} \,,\;(1, 0, \dots, 0, 0) \right)
+ \left(  {\bf 4} \,,\;(0, 0, \dots, 0, 1) \right)\;,
\nn
\ea
implying that the $T$-tensor consists of
\ba
T^{IJ,KL} &=& \left( 
{\bf 1} +{\bf 15}+{\bf 20'}\,,\; (0, 0, \dots, 0, 0) \right)
\;,
\non
T^{\Ga\Gb} &=& \left(  {\bf 1} \,,\; 2\cdot (0, 0, \dots, 0, 0)
+ 2\cdot (1, 0, \dots, 0, 1)+ (0, 1, \dots, 1, 0) 
\right)
\;,
\non
T^{IJ\Ga} &=& \left(  {\bf 15} \,,\; (0, 0, \dots, 0, 0)
+ (1, 0, \dots, 0, 1) \right)
\;,
\non
T^{AB} &=& \left(  {\bf 1} +{\bf 15} \,,\; (0, 0, \dots, 0, 0)
+ (1, 0, \dots, 0, 1) \right) \non
&&{}
+
\left(  {\bf 6} \,,\; (0, 1, \dots, 0, 0)+ (0, 0, \dots, 1, 0)
\right)
\;,
\non
T^{IJ,A} &=& \left(  {\bf \overline{4}} +{\bf 20} \,,\; 
(1, 0, \dots, 0, 0)\right) +
\left(  {\bf 4} +{\bf \overline{20}} \,,\;
(0, 0, \dots, 0, 1)\right)
\;,
\non
T^{\Ga A} &=& \left(  {\bf 4}\,,\; 
(1, 0, \dots, 1, 0) + 2\cdot (0, 0, \dots, 0, 1)\right) 
\;,
\non
&&{}+
\left(  {\bf \overline{4}}\,,\; 
(0, 1, \dots, 0, 1) + 2\cdot (1, 0, \dots, 0, 0)\right) 
\;,
\label{exT6}
\ea
with a linear relation between the four singlets, a linear relation
between the three representations in the $\left( {\bf 1} \,,\; (1, 0,
\dots, 0, 1)\right) $, $\left( {\bf 4} \,,\; (0, 0, \dots, 0,
1)\right) $, $\left( {\bf \overline{4}} \,,\; (1, 0, \dots, 0,
0)\right) $, and $\left( {\bf 15} \,,\; (0, 0, \dots, 0, 0)\right) $,
respectively, and where the two representations $\left( {\bf 15} \,,\;
(1, 0, \dots, 0, 1)\right) $ coincide.

\bigskip

\mathon
{\bf $N=5\,$}:$\;\;$
\mathoff
Under $\Sp{2}\times \Sp{k}$, the adjoint representation
of $\Sp{2,k}$ decomposes into
\ba
X^{IJ}\!: ~ \left(  {\bf 10} \,,\; (0, 0, 0,\dots ) \right)\;,
\quad
X^{\Ga}\!: ~ \left(  {\bf 1} \,,\; (2, 0, 0,\dots ) \right)\;,
\quad
Y^A\!:~  \left(  {\bf 4} \,,\; (1, 0, 0,\dots ) \right)\;,
\nn
\ea
implying that the $T$-tensor consists of
\ba
T^{IJ,KL} &=& \left( 
{\bf 1} +{\bf 5}+{\bf 14}\,,\; (0, 0, 0,\dots ) \right)
\;,
\non
T^{\Ga\Gb} &=& \left(  {\bf 1} \,,\; (0, 0, 0,\dots ) +(0, 1, 0,\dots )+
(0, 2, 0,\dots ) \right)
\;,
\non
T^{IJ\Ga} &=& \left(  {\bf 10}\,,\; (2, 0, 0,\dots )  \right)
\;,
\non
T^{AB} &=& \left( 
{\bf 1} + {\bf 5} \,,\; (0, 0, 0,\dots )+(0, 1, 0,\dots )  \right)
+ \left({\bf 10}  \,,\;  (2, 0, 0,\dots ) \right)
\;,
\non
T^{IJ,A} &=& \left(  {\bf 4} +  {\bf 16}\,,\; (1, 0, 0,\dots )  \right)
\;,
\non
T^{\Ga A} &=& \left(
{\bf 4} \,,\;  (1, 0, 0,\dots ) + (1, 1, 0,\dots ) \right)\;,
\label{exT5}
\ea
where the two representations $\left( {\bf 10} , (2, 0, 0,\dots )
\right)$ in $T^{IJ\Ga}$ and $T^{AB}$ coincide.

\end{appendix}

{\small

\providecommand{\href}[2]{#2}\begingroup\raggedright
\endgroup

}

\end{document}